\documentstyle[psfig,sprocl]{article}

\def\mstop{m_{\,\widetilde{t}}}
\def\mstopeff{m_{\,\widetilde{t}}^{\rm eff}}

\def\bold#1{\setbox0=\hbox{$#1$}%
     \kern-.025em\copy0\kern-\wd0
     \kern.05em\copy0\kern-\wd0
     \kern-.025em\raise.0433em\box0 }

\def\GENITEM#1;#2{\par\vskip6pt \hangafter=0 \hangindent=#1
   \Textindent{$ #2$ }\ignorespaces}

\def\be{\begin{equation}}
\def\ee{\end{equation}}
\def\bea{\begin{eqnarray}}
\def\eea{\end{eqnarray}}
\def\simlt{\stackrel{<}{{}_\sim}}
\def\simgt{\stackrel{>}{{}_\sim}}

\def\NPB#1#2#3{{\it Nucl.~Phys.} {\bf{B#1}} (19#2) #3}
\def\PLB#1#2#3{{\it Phys.~Lett.} {\bf{B#1}} (19#2) #3}
\def\PRD#1#2#3{{\it Phys.~Rev.} {\bf{D#1}} (19#2) #3}
\def\PRL#1#2#3{{\it Phys.~Rev.~Lett.} {\bf{#1}} (19#2) #3}
\def\ZPC#1#2#3{{\it Z.~Phys.} {\bf C#1} (19#2) #3}
\def\PTP#1#2#3{{\it Prog.~Theor.~Phys.} {\bf#1}  (19#2) #3}
\def\MPLA#1#2#3{{\it Mod.~Phys.~Lett.} {\bf#1} (19#2) #3}

\def\AP#1#2#3{{\it Ann.~Phys.} {\bf#1} (19#2) #3}

\def\HPA#1#2#3{{\it Helv.~Phys.~Acta} {\bf#1} (19#2) #3}
\def\JETPL#1#2#3{{\it JETP~Lett.} {\bf#1} (19#2) #3}

\def\mst11{m_{\;\widetilde{t}_{1}}}

\def\mst22{m_{\;\widetilde{t}_{2}}}
\def\mst12{m_{\;\widetilde{t}_{1,2}}}

\def\msb11{m_{\;\widetilde{b}_{1}}}
\def\msb22{m_{\;\widetilde{b}_{2}}}
\def\msb12{m_{\;\widetilde{b}_{1,2}}}

\def\mwidetilde2{\widetilde{m}^{2}}

\begin{document}
\pagestyle{empty}
\begin{flushright}
FERMILAB-Pub-97/95-T\\
CERN-TH/97-74 \\
hep-ph/9704347 
\end{flushright}
\vspace*{5mm}
\begin{center}
{\Large \bf Electroweak Baryogenesis and 
Higgs Physics~\footnote{To appear in {\it Perspectives on
Higgs Physics II}, ed. G.L. Kane, World Scientific, Singapore}}\\
\vspace{1.5cm}
{\large M. Carena$^{\dagger}$ and
C.E.M. Wagner$^\ddagger$}\\
\vspace{0.3cm}
$^\dagger$ Fermi National Accelarator Laboratory \\
P.O. Box 500, Batavia, IL 60510, USA\\
\vspace{0.3cm}
$^\ddagger$ Theory Division, CERN\\
CH-1211 Geneva 23, Switzerland\\
\vspace{0.3cm}
\vspace*{2cm}
Abstract
\end{center}
Electroweak Baryogenesis 
is a particularly attractive theoretical scenario, since it relies
on physics which can be tested at present high energy collider
facilities. Within the Standard Model,
it has been shown that
the requirement of preserving the  baryon number 
generated at the weak scale leads
to strong bounds on the Higgs mass, which are already inconsistent with
the present experimental limits. 
In the  Minimal Supersymmetric
extension of the Standard Model,  
we demonstrate that light stop effects can render 
the electroweak phase transition 
sufficiently strongly first order,
opening the possibility of electroweak baryogenesis
 for values of the Higgs mass
at the  LEP2 reach. The generation of the observed baryon asymmetry
also requires small chargino masses and new CP-violating
phases associated
with  the 
stop and Higgsino mass parameters. We discuss the direct
experimental tests of this 
scenario and other  relevant phenomenological
issues related to it.
\vfill
\begin{flushleft}
April 1997
\end{flushleft}
\eject
\pagestyle{empty}
\setcounter{page}{1}
\pagestyle{plain}
\newpage
\title{ELECTROWEAK BARYOGENESIS AND HIGGS
PHYSICS}
\author{M. CARENA},
\address{FERMILAB, P.O. Box 500, Batavia, IL
60510, USA}
\author{C.E.M. WAGNER}
\address{CERN, TH Division, CH--1211 Geneva 23, Switzerland}
\maketitle\abstracts{
Electroweak Baryogenesis 
is a particularly attractive theoretical scenario, since it relies
on physics which can be tested at present high energy collider
facilities. Within the Standard Model,
it has been shown that
the requirement of preserving the  baryon number 
generated at the weak scale leads
to strong bounds on the Higgs mass, which are already inconsistent with
the present experimental limits. 
In the  Minimal Supersymmetric
extension of the Standard Model  
we demonstrate that light stop effects can render 
the electroweak phase transition 
sufficiently strongly first order,
opening the possibility of electroweak baryogenesis
 for values of the Higgs mass
at the  LEP2 reach. The generation of the observed baryon asymmetry
also requires small chargino masses and new CP-violating
phases associated
with  the 
stop and Higgsino mass parameters. We discuss the direct
experimental tests of this 
scenario and other  relevant phenomenological
issues related to it.}
%
%
\section{INTRODUCTION}

One of the fundamental problems of particle physics is to
understand  the origin of the observed baryon asymmetry of
the universe.  The mechanism for the generation of baryon number
may rely on physics at very high energies, of order of the
Planck or Grand Unification scale and hence difficult to test at
present high energy colliders. Even in this case,
the final result for the baryon
asymmetry will always be affected by low energy physics. Indeed,
although baryon number is preserved  in the Standard Model 
at the classical level, it
is violated through anomalous processes at the
quantum level \cite{anomaly}.
As the Universe cools down,  unless 
specific conditions on the baryon and lepton asymmetries generated
at high energies are fulfilled \cite{HerbiG}, 
the anomalous 
processes will tend to erase  the baryon asymmetry generated
at high energies. If the baryon asymmetry is
completely washed out at temperatures far above the weak scale,
the observed baryon number must proceed from physical processes
at temperatures close to $T_c$, at which
the electroweak phase transition takes place.

Baryogenesis at the electroweak phase transition is a very
attractive alternative since it relies only on physics at
the weak scale, and it is hence testable in the near future.
In principle, the
Standard Model (SM) fulfills all the
requirements~\cite{baryogenesis} for a successful generation
of baryon number~\cite{reviews}. Non-equilibrium processes
occur at the first order electroweak phase transition, baryon
number is violated by anomalous processes and CP is violated
by explicit phases in the CKM matrix. In order to quantitatively
estimate the generated baryon number, one should take into
account that the baryon number violating 
processes are effective also after the electroweak phase
transition, and are only suppressed by a Boltzman factor
\begin{equation}
\Gamma \simeq T^4  \exp\left(-\frac{E_{\rm sph}}{T} \right),
\label{rate}
\end{equation}
where the sphaleron energy,
$E_{\rm sph}$, is equal to the height of the barrier separating
two topologically inequivalent vacua \cite{sphalerons}. The sphaleron
energy is given by
\begin{equation}
E_{\rm sph}(T) \simeq B \frac{2 M_W(T)}{\alpha_w(T)}
\end{equation}
with $B \simeq {\cal O}(2)$ being  a slowly
varying function of the Higgs quartic coupling, 
$M_W$ the weak gauge boson mass and
$\alpha_w$ the weak gauge coupling.
If the phase transition were second order, or very weakly first
order, the baryon number violating processes would be approximately
in equilibrium and no effective
baryon number will survive at $T \simlt T_c$.
Comparing the rate, Eq. (\ref{rate}), 
with the rate of expansion of the universe
one can obtain the condition under which the generated
baryon number will be preserved after the 
electroweak phase transition. This implies a bound on the
sphaleron energy,
$E_{\rm sph}(T_c)/T_c \simgt 45$ \cite{first} or, equivalently,
\begin{equation}
\frac{v(T_c)}{T_c} \simgt 1. \;\;\;\;\;
\label{orderp}
\end{equation}

Since $v(T_c)/T_c$ is inversely proportional to the quartic
coupling appearing in the low
energy Higgs effective potential, the requirement
of preserving  the generated
baryon asymmetry puts an upper bound on
the value of the
Higgs mass. The actual bound  depends on the
particle content of the theory
at energies of the order of the weak scale.
In the case of the Standard Model, the present experimental bounds
on the  Higgs mass are already too strong, rendering
the electroweak phase transition  too weakly first order.
Hence, within the Standard Model, the generated baryon asymmetry at
the electroweak phase transition cannot be preserved~\cite{first},
as perturbative~\cite{improvement}$^-$\cite{pertres2} and
non-perturbative~\cite{nonpert,Jansen} analyses have shown. It is
interesting to notice that, even in the absence of experimental
bounds, the requirement of a sufficiently strong first order
phase transition leads to bounds on the Higgs mass such that
the electroweak breaking minimum would become unstable unless
new physics were present at scales of the order of the weak
scale \cite{earlyst}$^-$\cite{CEQ}. 
We shall review these bounds below.
On the other hand,
CP-violating processes are suppressed by powers of $m_f/M_V$,
where $m_f$ are the light-quark masses and $M_V$ is the mass of
the vector bosons. These suppression factors are sufficiently strong
to severely restrict the possible baryon number
generation~\cite{fs,huet}.
Therefore, if the baryon asymmetry is generated at the electroweak phase
transition,
it will require the presence of new physics at the electroweak scale.

The most natural extension of the Standard Model, which naturally
leads to
small values of the Higgs masses is low energy supersymmetry.
It is hence highly interesting to test under which conditions baryogenesis
is viable within this 
framework ~\cite{early}$^-$\cite{mariano2}.
It was recently shown ~\cite{CQW}$^-$\cite{Delepine}
that the phase transition can be sufficiently strongly first
order only in a restricted region of parameter space, which strongly
constrains the possible values of the lightest
stop mass, 
of the lightest CP-even Higgs mass (which should be at the reach
of LEP2) and of  the ratio of vacuum expectation
values, $\tan\beta$.
These results have been confirmed by explicit sphaleron
calculations
in the Minimal Supersymmetric Standard Model
(MSSM)~\cite{MOQ}.

On the other hand, the Minimal Supersymmetric Standard Model,
contains, on top of the CKM matrix phase, additional sources of CP-
violation
and can account for the observed baryon asymmetry.\footnote{An
interesting scenario, relying only on the Standard Model phases
was recently suggested \cite{Worah}. However,
since it requires a large mixing
between the second and third generation squarks, an analysis of the
strength of the first order phase transition will be necessary to decide
about its viability.}
New CP-violating phases can arise
from the soft supersymmetry breaking parameters associated with
the  stop mixing angle.  Large values of the mixing angle
are, however, strongly restricted in order to preserve a
sufficiently strong first
order electroweak phase
transition ~\cite{mariano1}$^-$\cite{CQW}.
Therefore, an acceptable baryon asymmetry
requires  a delicate balance between the
value of the different mass parameters
contributing
to the left-right stop mixing, and their associated CP-violating
phases. Moreover, the CP-violating currents must originate from
the variation of the CP-odd phases appearing in the couplings of
stops, charginos and neutralinos to the Higgs particles. This
variation would be zero if $\tan\beta$ were a constant,
implying that the heavy Higgs doublet can not decouple at scales
far above $T_c$, or equivalently, the CP-odd Higgs mass should
not be much larger than $M_Z$.  On the other hand, since
the phase transition
becomes weaker for  lighter CP-odd Higgs bosons, a restricted
range for the CP-odd and charged Higgs masses may be obtained
from these considerations.

The scenario of Electroweak Baryogenesis (EWB) has crucial implications
for Higgs physics and imposes important constraints on the
supersymmetry breaking parameters. Most appealing, this scenario
makes definite predictions, which may be tested at present or
near future colliders.

In section 2 we shall present the improved one-loop finite
temperature Higgs effective potential. In section 3 we discuss
the Standard Model case, on the light of present experimental
constraints on the Higgs mass
and the requirement of stability of the 
physical vacuum. In section 4, we study the 
strength of the electroweak phase transition
within the minimal supersymmetric
extension of the standard model, discussing in detail the effect
of light stops in expanding the allowed Higgs mass range 
and analyzing the conditions to avoid color breaking minima.
We also discuss the strength of the electroweak phase transition
in the cases of large and small values of the CP-odd Higgs mass, and 
analyse
the new sources of CP-violation which may contribute 
to the generation of baryon asymmetry within the
MSSM. In section 5 we study the generation of baryon asymmetry. 
In section 6 we ellaborate on the experimental tests of this
scenario both at LEP2 and the Tevatron, and discuss the predictions
for some rare flavor changing neutral current processes within this
framework. 
In section 7 we summarize
the prospects for electroweak baryogenesis.

\section{FINITE TEMPERATURE HIGGS EFFECTIVE POTENTIAL}

As we explained above, the requirement of preserving  the
baryon asymmetry after the phase transition implies that $v(T_c)/T_c$
must be larger than one.  To extract the implications
of this requirement, a detailed knowledge of the
finite temperature effective potential of the Higgs field
is needed. The finite
temperature effective potential for the neutral component of the
Higgs field may be  computed at the
one-loop level \cite{DJ},
\begin{eqnarray}
V(\phi,T) = V_{\rm tree}(\phi) + V_1(\phi,0) + V_1(\phi,T)
\end{eqnarray}
where $V_{\rm tree}(\phi)$, $V_1(\phi,0)$ and $V_1(\phi,T)$ are the
tree level, one-loop zero temperature and one-loop finite temperature
contributions to the effective potential, respectively. 
Their expressions are given
by
\begin{eqnarray}
V_{\rm tree}(\phi)& = &m^2 \phi^2 + \frac{\lambda}{2} \phi^4
\nonumber\\
V_1(\phi,0) & = & \sum_i \frac{n_i}{64 \pi^2}
m_i^4(\phi) \left[ \log\left(
\frac{m_i^2(\phi)}{Q^2}\right) - c_i \right]
\nonumber\\
V_1(\phi,T) & = & \sum_i \frac{n_i}{2 \pi^2 \beta^4}
\int_0^{\infty} dx \; x^2 \log\left( 1 \pm \exp-(x^2 +
\beta^2 m_i^2(\phi))^{1/2}
\right),
\label{effpot}
\end{eqnarray}
where $m_i(\phi)$ is the mass of the $i$-field in the
background of the field $\phi$, $n_i$ is its total number of
degrees of freedom, $c_i = 5/6$ for vector bosons and 3/2
for scalars and fermions.
$\beta^{-1}$ is proportional to the
temperature and the plus and minus sign in $V_1(\phi,T)$
are associated with
fermionic and bosonic particles, respectively.
Observe that the contribution of heavy particles to
the temperature dependent part of the effective potential is
exponentially suppressed.
For  values of the masses $m_i(\phi) \simlt 2 \; T$,
the effective potential admits a high temperature expansion.
In this limit,
the contribution of bosonic particles 
to the Higgs effective potential is given by~\cite{DJ}
\begin{equation}
V_1^{\rm b}(\phi,T) =
\sum_i n_i \left\{ \frac{m_i^2(\phi)}{24 \beta^2} - \frac{1}{12 \pi}
\frac{m_i^3(\phi)}{\beta} - \frac{1}{64 \pi^2} m_i^4(\phi)
\log(m_i^2(\phi) \beta^2) +... \right\}
\label{eq:cubic}
\end{equation}
while that of fermions is given by
\begin{equation}
V_1^{\rm f}(\phi,T)  =
\sum_i n_i \left\{ \frac{m_i^2(\phi)}{ 48 \beta^2} +
\frac{m_i^4(\phi)}{64 \pi^2} 
\log\left(m_i^2(\phi) \beta^2\right) +...\right\}.
\end{equation}
Therefore,
in the region of validity of the high temperature expansion, the
effective potential reads,
\begin{equation}
V(\phi,T) = D (T^2 - T_0^2) \phi^2
- E T \phi^3 + \frac{\lambda}{2} \phi^4 +...
\end{equation}
where $D$, $E$ and $\lambda$ are temperature dependent functions,
$T_0$ is the temperature at which the curvature of the potential
vanishes at the origin and
we have chosen the normalization such that $<\phi> = v/\sqrt{2}$,
with $v(0) \sim 246$ GeV.   The
minimization of the potential at $T = T_c$, the temperature at
which the electroweak symmetry breaking  and
the electroweak symmetry preserving
minima become degenerate,  leads to
\begin{equation}
\frac{\phi(T_c)}{T_c} = \frac{E}{\lambda}.
\label{eq:vot}
\end{equation}
Hence, the strength of the phase transition is directly proportional
to the coefficient of the cubic term induced by the presence of
bosonic particles, like the gauge bosons,
with masses $m_B = k_B \phi^2$ (see Eq. (\ref{eq:cubic})).

Higher loop corrections are important to define the correct
infrared properties of the effective potential, and to reduce its
gauge dependence. The most important corrections
come from the so-called Daisy graphs \cite{improvement}, 
which effectively amount
to replace
\begin{equation}
\sum_i \frac{n_i \; m_i^3(\phi)}{12 \pi} \rightarrow
\sum_i \frac{n_i \; m_i^3(\phi,T)}{12 \pi}
\end{equation}
in Eq. (\ref{eq:cubic}), 
where
\begin{equation}
m_i^2(\phi,T) = m_i^2(\phi) + \Pi(T)
\label{eq:thermalm}
\end{equation}
and
$\Pi(T)$ is the temperature dependent vacuum polarization
contribution to the thermal masses.

An important observation is that the strength of the
phase transition is inversely proportional to the squared
of the Higgs mass. This is due to the fact that, at zero
temperature
\begin{equation}
m_H^2 = \lambda  \; v^2,
\label{mhiggs}
\end{equation}
and the value of $\lambda$ at the critical temperature is of
the order of its zero temperature value. Hence, 
from Eqs. (\ref{orderp}), (\ref{eq:vot}) and (\ref{mhiggs}),
the requirement
of preservation of the baryon asymmetry leads to an upper bound
on the Higgs mass. 

\section{THE STANDARD MODEL CASE}
\subsection{The Electroweak Phase Transtion}

In the  Standard
Model,  the effect of including thermal masses, 
Eq. (\ref{eq:thermalm}),
implies that
only the transverse  modes of the electroweak gauge
bosons will contribute to the cubic term in the effective
potential and hence the Daisy improvement
leads to a weaker phase
transition 
than the one predicted at the one-loop level.
The coefficient of the cubic term is
given by
\begin{equation}
E_{\rm SM}  = \frac{2}{3} \left(\frac{2 M_w^3 + M_Z^3}{\sqrt{2} \pi v^3}
\right)
\end{equation}
and hence
the upper bound on the Higgs mass derived from Eqs. (\ref{orderp}) and 
(\ref{eq:vot}) is
\begin{equation}
m_H \simlt 40 \; {\rm GeV}.
\label{eq:bound1l}
\end{equation}

Although the Daisy resummation includes 
the dominant higher loop corrections
to the effective potential, there are additional
two-loop effects which have been shown to lead to
non-negligible corrections \cite{twoloop}, making the
phase transition more strongly first order and
increasing slightly the above upper bound on the Higgs mass.
Non-perturbative effects have been taken into account through
lattice studies \cite{nonpert}.
These simulations have been done both
in four dimensions as in the effective three dimensional 
 theory arising at high  temperatures.
A more involved perturbative resummation has been performed~\cite{pertres2},
 showing
excellent agreement with the lattice results \cite{Jansen}.
In general, the results  for the upper
bound on the Higgs mass derived from the 
non-perturbative studies
do not differ in a significant way from those 
ones coming from perturbative analyses.
Numerically, the upper bound obtained from the lattice is
somewhat higher than the results obtained from
the improved one-loop analysis. 
The result of the one-loop analysis,
Eq. (\ref{eq:bound1l}),
may be hence
quoted as a conservative bound on the Higgs mass.

\subsection{Stability Bounds and Experimental Limits on $m_H$.}

The low values of the Higgs mass required to  preserve
the baryon asymmetry are clearly in  conflict with the
current experimental bounds on this quantity. The present LEP bound
on the Standard Model Higgs mass reads \cite{Alephl}
\begin{equation}
m^{SM \; exp.}_H \simgt 70 {\rm GeV}
\end{equation}
and hence, for the mechanism of electroweak baryogenesis to
survive, new physics
should be present at the weak scale.

Actually, this argument might have been made even in the absence
of experimental constraints, by analysing the stability of
the physical vacuum~\cite{earlyst}$^-$\cite{CEQ}.
This may be understood by considering
the renormalization group
improved  effective potential for the neutral Higgs
at zero temperature, which is approximately given by
\begin{equation}
V(\phi)   = m^2 \phi^2  + \frac{\lambda(\phi)}{2} \phi^4,
\label{eq:Veff}
\end{equation}
where $\lambda(\phi)$ means that the quartic coupling must be
evaluated at the relevant scale at which the
effective potential is analysed.  
The dominant contributions to
the renormalization
group equation of the Higgs quartic coupling are
\begin{equation}
\frac{d \lambda}{dt} = \frac{3}{8 \pi^2} \left( \lambda^2
+ \lambda h_t^2 - h_t^4 \right) + {\rm electroweak} \;
{\rm corrections},
\label{eq:lambda}
\end{equation}
where $h_t$ is the top quark Yukawa coupling,
$t = \log(Q^2/\Lambda^2)$, with $Q$  the renormalization group
scale and $\Lambda$  the cutoff of the effective theory. 
For large values of $\lambda$, the quartic coupling of the
Higgs fields, grows indefinitely with rising energy and
an upper  bound on $m_H$ follows from the requirement 
of perturbative consistency of the theory
up to a given cutoff scale $\Lambda$ below $M_{\rm Pl}$.
The upper bound on $m_H$ depends mildly on the top quark
mass through the impact of the top quark Yukawa coupling
on the running of the quartic coupling $\lambda$.

On the other hand, the effect of the  large values of $h_t$
on the renormalization group evolution of the quartic 
coupling, may drive $\lambda$ to negative values at large
energy scales, thus destabilizing the standard electroweak
vacuum. The requirement of vacuum stability in the Standard
Model imposes a lower bound on the Higgs boson mass for a
given cutoff scale. This bound on $m_H$ is defined as the lower
value of $m_H$ for which $\lambda(\phi) \geq 0$ for any value
of $\phi$ below the scale $\Lambda$ at which new physics beyond
the Standard Model should appear. From Eq. (\ref{eq:lambda}),
it is clear that the stability condition of the effective 
potential demands new physics at lower scales for larger
values of $m_t$ and lower values of $m_H$.

Fig. 1 \cite{LEPRep} shows the
perturbativity and stability bounds on $m_H$ 
as a function of the physical top quark mass $M_t$, for different
values of the cutoff $\Lambda$ at which new physics is
expected. 
Present experimental results lead to a precise knowledge of the
value of the top quark mass, $M_t = 175 \pm 6$ GeV \cite{TopM}. 
Hence, as follows from Fig. 1, independently
of the experimental bounds on the Higgs mass, the theoretical
upper bound on the Higgs mass obtained from the requirement of
preserving  the baryon asymmetry, $m_H \simeq 40$ GeV, 
implies an upper bound on
the scale of new physics of the order of the electroweak scale. 
\begin{figure}
\vspace{-1 cm}
\centerline{
\psfig{figure=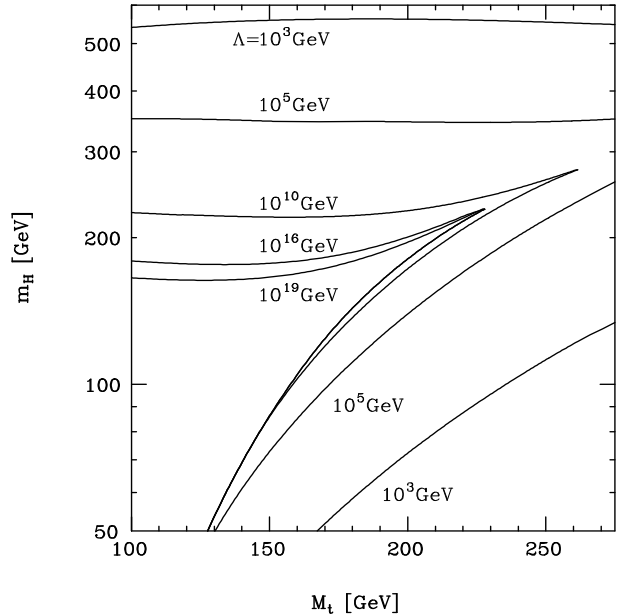,width=14cm,height=10cm,angle=90}}
\caption[]{{\it Bounds on the Higgs mass as a function
of the top quark mass for different values of the scale
$\Lambda$, at which new physics is expected to
appear.}}
\label{fig:1}
\end{figure}
This new
physics will affect the structure of the effective potential
at the weak scale, and hence the upper bound on the Higgs mass
derived from requiring a sufficiently strong first order
phase transition  has to be revised.

\section{BEYOND THE STANDARD MODEL: SUPERSYMMETRY}

The arguments presented in  section 3 depend
strongly on the structure of the effective Higgs potential.
Hence, the
Higgs mass bounds could be simply avoided by complicating the
Higgs structure. Models with two Higgs doublets are among the
simplest ones, and hence they have attracted some attention.
Two Higgs doublet models, in general, lead to charge breaking
minima, unless the vacuum expectation values of 
both Higgs doublets are alligned in
such a way that the electromagnetic symmetry is preserved.
Moreover, they generally lead to flavor changing neutral
currents which are beyond the present experimental limits.
There are several ways to avoid these problems,
and models of this type have been analysed in the literature.
However, there is no clear motivation for the extension
of the Higgs sector within the
Standard Model. On the contrary,
two Higgs doublets are necessary  in the context
of supersymmetric theories.

The most appealing extension of the Standard Model is the Minimal
Supersymmetric Standard Model (MSSM) \cite{reviewsu}.
Supersymmetry relates bosonic and
fermionic degrees of freedom. For each chiral fermion (gauge boson)
of the Standard Model, a complex scalar (Majorana fermion) appears
in the theory, with equal gauge quantum numbers as the Standard
Model field ones. Moreover, supersymmetry implies a relation between
the couplings of the bosonic  and fermionic degrees of freedom,
yielding a cancellation of the
quadratic divergencies associated with the radiative
corrections to the scalar Higgs
masses, and providing
a technical explanation of the hierarchy stability from $M_{Pl}$ to the
electroweak scale.

Supersymmetry provides a solution to most of the
problems affecting the multi-Higgs systems.
Two Higgs doublets are naturally required,  to
cancel  the anomalies generated
by the superpartners of the Higgs bosons. Moreover, flavor changing
neutral currents are
naturally suppressed since supersymmetry
requires that only one of the Higgs doublets couples to
the up-like (down-like) quarks.
In addition, the effective Higgs potential is
such that the vacuum state is naturally alligned towards a charge
conserving minimum in the Higgs sector of these models.

The Higgs spectrum of the Minimal Supersymmetric extension of
the Standard Model
consists of two CP-even bosons, a CP-odd
and a charged Higgs bosons \cite{Hhunter}. 
The heaviest CP-even and the charged
Higgs masses are of the order of the CP-odd Higgs mass, $m_A$,
and for large values of $m_A$ they form a heavy Higgs doublet
which decouples at low energies. In this limit,
the lightest Higgs doublet
contains the three Goldstone modes, as well as a CP-even state.
Moreover, supersymmetry relates the Higgs quartic couplings to the
weak gauge couplings leading to
an upper bound on the lightest CP-even
Higgs mass,
\begin{equation}
m_h^2 \leq M_Z^2 \cos^2 2\beta + {\rm rad.} \;
{\rm  corr.},
\label{eq:htree}
\end{equation}
where $\tan\beta = v_2/v_1$ and  $v_2$ ($v_1$) is  the
vacuum expectation value of the Higgs field $H_2$ ($H_1$)
which couples to the up (down) quarks.
The last term in the above equation denotes the
radiative corrections, which are induced through supersymmetry
breaking effects. The main contributions
will be discussed in section 4.1.

Supersymmetry is particularly appealing for 
the scenario of electroweak
baryogenesis, since it naturally provides
small values of
the Higgs mass, and hence tends to give a relatively strong first
order phase transition. Moreover, in a supersymmetric theory
the negative contributions of the top quark to the 
renormalization group evolution of  the Higgs quartic couplings 
are
compensated by the effects of its supersymmetric partner, providing
a natural solution to the vacuum stability problem. However,
since supersymmetry is broken in nature, this argument depends strongly
on the supersymmetry breaking scale. Indeed the supersymmetry
breaking scale may be identified with the scale of new 
physics (see section 3).
Since the  particles which couple more strongly to the
Higgs are the top quark and its supersymmetric partners, the relevant
scale of new physics, in relation to the
stability of the Higgs potential,
is given by  the stop masses. From Fig. 1 we see
that  in order to
preserve the stability of the effective potential, for
a Higgs mass of  order of the present experimental
bound, the lightest stop mass must be of the order of
the weak scale.

\subsection{Higgs and Stop Masses in the MSSM}

The stop masses have an incidence on the Higgs potential which
goes beyond the problem of vacuum stability. The stop radiative
corrections affect the value of the parameters appearing in the
effective potential of the Higgs field in a way which 
depends on
the exact value of the stop masses \cite{Higgs1l}. 
For values of the stop masses
close to the top mass, there is an approximate cancellation
between the top and stop  loop effects and the tree-level
relation between $m_h$ and $M_Z$, Eq. (\ref{eq:htree}),
is recovered. For very large
values of the stop masses, instead, the tree-level relation is
strongly affected by radiative corrections and the effective
theory is similar to a non-supersymmetric two Higgs doublet model.

It is  interesting
to discuss in some detail the properties of the superpartners of
the top quark. The left handed and right
handed stops are not mass eigenstates, due to the appearence of
effective trilinear couplings between the left- and right-handed
stops and the Higgs fields
\begin{equation}
{\cal L}_{3} = - h_t \left( \epsilon_{ij} A_t H_2^j Q^i U
- \mu^* H_1^{* i} Q^i U  \right) + h.c.,
\end{equation}
where $Q$ is the scalar
top-bottom left-handed doublet and
$U$ is the charge conjugate of the right handed
scalar top,
$A_t$ is a soft supersymmetry
breaking mass parameter and $\mu$ is the Higgs superpartner
(Higgsino) mass term. Once the neutral components of the
Higgs doublets acquire vacuum expectation
values, a mixing term appears between the left and right handed stops,
leading to the following mass matrix
\begin{equation}
{\cal M}_{st}^2 = \left[
\begin{array}{cc}
m_Q^2 + m_t^2 + D_L  &  m_t \left(A_t - \mu^*/\tan\beta\right)  \\
m_t \left(A_t^* - \mu/\tan\beta\right)
&  m_U^2 + m_t^2 +D_R
\end{array} 
\right]
\equiv
\left[
\begin{array}{cc}
m^2_{LL}  &  m^2_{LR} \\
m^2_{LR}  &  m^2_{RR}  
\end{array} 
\right] \;\;,
\label{eq:stopmatrix}
\end{equation}
where $m_Q^2$ and $m_U^2$ are the soft supersymmetry breaking
square mass
parameters of the left and right handed stops, respectively,
$D_L$ and $D_R$ are the (relatively small) D-term contributions
to the stop masses, and $m_t = h_t <H_2>$ is the running
top quark mass.
The stop mass eigenvalues are then given by
\begin{equation}
m^2_{\widetilde{T},\tilde{t}}= \frac{m^2_{LL} + m^2_{RR}}{2} \pm
\sqrt{\left(\frac{m^2_{LL} - m^2_{RR}}{2}\right)^2 +|m^2_{LR}|^2}.
\end{equation} 

The lightest CP-even Higgs mass is a monotonically increasing
function of the CP-odd Higgs mass $m_A$. 
As mentioned above, for large values
of the CP-odd Higgs mass,  $m_A \gg M_Z$,
the heavy Higgs doublet
decouples and we obtain an upper bound on the
lightest CP-even Higgs mass. This value has been computed
at the one and two-loop level, and considering a renormalization
group resummation \cite{Higgs1l}. In the large $m_A$ limit, with
$m_Q \simeq m_U$, a simple formula is obtained at the two-loop
level \cite{CEQW},
\begin{eqnarray}
m_h^2 & = & M_Z^2 \cos^2 2 \beta \left(1 - \frac{3 m_t^2}{4 \pi^2
v^2} t \right)
\nonumber\\
& + & \frac{3m_t^4}{2\pi^2v^2} \left[ \frac{X_t}{2}
+ t + \frac{1}{16\pi^2} \left(
\frac{3 m_t^2}{v^2} - 32 \pi \alpha_3\right)
\left( X_t t + t^2 \right) \right]
\label{Higgs2l}
\end{eqnarray}
where $m_t$  and $\alpha_3 = g_3^2/4 \pi$, with $g_3$ the strong gauge 
coupling, are evaluated
at the top quark
mass scale,
\begin{equation}
t = \log\left(\frac{M_S^2}{m_t^2}\right)
\;\;\;\;\;\;\;\;\;\;\;\;
X_t = \frac{2 |\tilde{A}_t|^2}{M_S^2} \left(
1 - \frac{\left|\tilde{A}_t\right|^2}{12 M_S^2} \right)
\end{equation}
with $\tilde{A}_t = A_t - \mu^{*}/\tan\beta$,
$M_S^2 = (m_Q^2 + m_U^2)/2 + m_t^2$, and we have
implicitly assumed that $|m_t \tilde{A}_t| \simlt 0.5 M_S^2$.

The first term in Eq. (\ref{Higgs2l}) reproduces the tree-level
contribution to the lightest Higgs mass, Eq. (\ref{eq:htree}).
The tree level value
of the lightest Higgs mass increases for larger values of $\tan\beta$
and tends to zero for $\tan\beta$ equal to one. The most important
radiative corrections to the Higgs mass value are positive,
proportional to $m_t^4$, 
and increase logarithmically with the supersymmetry
breaking scale $M_S$.
The stop mixing parameter plays also a very important role in determining
the Higgs mass value. A maximum value for the Higgs mass is obtained
for $|\tilde{A}_t| \simeq \sqrt{6} M_S$. Such large values
of the stop mass mixing parameter are, however, disfavor by model
building considerations. As we shall show in section 4.2,
large values of $|\tilde{A}_t| \simgt 0.5 M_S$ also
make the phase transition more weakly first order.

Since the phase transition becomes stronger for lower values of the Higgs
mass, it is interesting to analyse the conditions under which the
lightest CP-even Higgs mass becomes close to the present experimental
bound $m_h \simgt 70$ GeV.
Low values of
the Higgs mass $m_h \simlt 85$ GeV are only obtained for
$\tan\beta \simlt$~4. Very low values of $\tan\beta$ are associated
with large values of the top quark Yukawa coupling, and 
for a given value of $m_t$ a lower bound
on $\tan\beta$ may be obtained by requiring perturbative consistency
of the theory up to scales of order of the grand unification scale.
This requirement leads to values of $\tan\beta \simgt 1.2$ for the
acceptable experimental range for the top quark mass \cite{IR}.  
If $\tan\beta$
is close to one, at least one of the stop masses has to be large, in
order to overcome the present experimental limits
on $m_h$. For large splittings
between the two stop mass eigenvalues, one has to go beyond 
the approximation of
Eq. (\ref{Higgs2l}) \cite{CEQW}. We shall
briefly discuss this case in section 4.2.

\subsection{The Electroweak Phase Transition}
\subsubsection{Lightest stop mass effects on  the phase transition.}

As we discussed in section 3, for a fixed Higgs mass,
the strength of the first order phase transition may be enhanced
by increasing the value of the effective cubic term in the finite
temperature Higgs potential. This may be achieved by the presence
of extra bosonic degrees of freedom~\cite{AndH}, 
with sizeable couplings to
the Higgs sector. Within the minimal supersymmetric model,
the  bosonic particles which couple strongly to the
Higgs which acquires vacuum expectation value are the
supersymmetric partners of the top quark. Since the cubic
term is screened by field independent mass contributions, a
relevant enhancement of the cubic term
of the effective Higgs potential
may only be obtained for small values of  the lightest stop mass 
$m_{\tilde t} \simlt m_t$ \cite{CQW}.

The lightest stop must be mainly right-handed 
in order to naturally
suppress its contribution 
to the parameter $\Delta\rho$ and hence
preserve a good agreement with the precision measurements at LEP.
This can be naturally achieved if the 
soft supersymmetry breaking mass parameter of the left-handed stop,
 $m_Q$,
is much larger than $M_Z$.
We shall first discuss the large 
CP-odd mass limit, $m_A \gg M_Z$. In this case, the
heaviest Higgs doublet decouples and the low
energy theory contains only one light Higgs boson, $\phi$, with similar
properties to the Standard Model one,
\begin{equation}
\phi = \cos\beta H_1 + \sin\beta H_2.
\end{equation}

For moderate left-right stop mixing,
from Eq. (\ref{eq:stopmatrix}) it follows that 
the lightest stop mass is then approximately given by
\be
\mstop^2  \simeq m_U^2 + D_R + m_t^2(\phi) \left( 1  -
\frac{\left|\widetilde{A}_t\right|^2}{m_Q^2}
\right)
\label{eq:lightst}
\ee
where $m_t(\phi) = h_t \sin\beta \; \phi$.
Hence,
the lightest stop contribution 
to the finite temperature Higgs potential,
necessary to overcome the
SM constraints on the Higggs mass, strongly depends
on the value of $m_U^2$.
At finite temperature, however, the field mass receives
vacuum polarization contributions which have a strong
impact on the size of the induced cubic terms in the
effective finite temperature Higgs potential. 
Indeed, 
\begin{equation}
m_{\tilde{t}}^2(\phi,T) =
m_{\tilde{t}}^2(\phi,0) + \Pi_R(T)
\end{equation} 
where $\Pi_R(T) \simeq 4 g_3^2
T^2/9
+h_t^2/6[1+\sin^2\beta\left(1-
|\widetilde{A}_t|^2/m_Q^2\right)]T^2$
is the finite
temperature
self-energy contribution to the right-handed
squarks~\cite{CQW,CE}.

The improved one-loop finite temperature effective potential
is then given by
\begin{equation}
V_{\rm eff}^{\rm MSSM} = m^2(T) \phi^2  - 
T \left[ E_{\rm SM} \phi^3 + (2 N_c) 
\frac{m_{\tilde{t}}^{3/2}(\phi,T)}{12 \pi} \right]
+ \frac{\lambda(T)}{2} \phi^4 +...
\label{eq:temppot}
\end{equation}
where $N_c = 3$ is the number of colors, and $E_{\rm SM}$ is the
strength of the cubic term in the Standard Model case. 

Observe that the heaviest stop leads to a relevant contribution to
the zero-temperature effective potential, which can be absorved
into a redefinition of the mass and quartic coupling parameters. 
Large values of $m_Q$ have the effect of increasing the value
of the Higgs mass. Indeed, for $m^2_Q \gg m^2_U$ and moderate
values of $\tilde{A}_t$, the lightest Higgs mass expression 
at the one-loop level reads,
\begin{equation}
m_h^2 = M_Z^2 \cos^2 2\beta + \frac{3m_t^4}{4 \pi^2 v^2}
\log\left(\frac{m_{\tilde{t}}^2 m_{\tilde{T}}^2}{m_t^4}
\right) + {\cal O}\left(\frac{\tilde{A}_t^2}{m_Q^2}\right)
\end{equation}
where $m_{\tilde{T}}^2 \simeq m_Q^2 + m_t^2$ 
is the  heaviest stop square  
mass~\footnote{The two loop corrections to the
Higgs mass, in the limit of $m_Q^2 \gg m_U^2$ 
are also available~\cite{CEQW}}. 
Although larger values of 
$m_h$ are welcome in order to avoid its experimental bound,
since they are associated with an increase of the quartic
coupling, they
necessarily lead to a weakening of the first order 
phase transition. Therefore, very large values of $m_Q$, above
a few TeV, are disfavored from this point of view. The finite
temperature effects of the heaviest stop are, instead,
exponentially suppressed.

From Eq. (\ref{eq:temppot})
it follows that, in general, as happens with the longitudinal 
components of the gauge bosons, the lightest stop contribution
to the effective potential does not induce a cubic term. This
is mainly due to the fact that  the  effective stop
plasma mass squared at $\phi =0$,
\begin{equation}
(m^{\rm eff}_{\;\widetilde{t}})^2 = -\widetilde{m}_U^2 + \Pi_R(T)
\label{plasm}
\end{equation}
with $\widetilde{m}_U^2 \equiv 
- m_U^2$, is generally positive and
of order of $T^2$. 
If the right handed stop plasma mass at $\phi = 0$, Eq. (\ref{plasm}),
vanished, a large
value of the effective cubic term would be obtained. Since
$v(T_c)/T_c \simeq \sqrt{2} E/\lambda$, 
an upper bound on $v(T_c)/T_c$ may be
obtained from these considerations, namely
\be
\frac{v(T_c)}{T_c} < \left(\frac{v(T_c)}{T_c}\right)_{\rm SM}
+ \frac{2 \; m_t^3  \left(1 -
|\widetilde{A}_t|^2/m_Q^2\right)^{3/2}}{ \pi \; v \; m_h^2}\ ,
\label{totalE}
\ee
where $m_t = \overline{m}_t(m_t)$ is the on-shell running top
quark
mass in the $\overline{{\rm MS}}$ scheme.
The first term on the right hand side of
Eq.~(\ref{totalE}) is the SM contribution
\be
\left(\frac{v(T_c)}{T_c}\right)_{\rm SM}
\simeq \left(\frac{40}{m_h[{\rm GeV}]}\right)^2,
\ee
and the second term is
the contribution that would be obtained through the right handed
stops in the limit of a vanishing plasma mass.
The upper bound on $v(T_c)/T_c$ is almost an order of magnitude
larger than the one obtained in the Standard Model, implying that
Higgs masses of order $M_Z$ may be consistent with electroweak
baryogenesis.
Although the
exact cancellation of the effective stop mass at $\phi = 0$
is not likely to occur, it is clear that a partial cancellation 
is necessary to increase the cubic term 
coefficient considerably~\footnote{Higher loop 
corrections are important, making
values of $m_U \simgt 0$ possible~\cite{JoseR}}. 
A small plasma mass can only be obtained through
sizeable values of $\widetilde{m}_U$, this means, negative sizeable values
of the right-handed stop mass parameter.  
Moreover, as it is clear from Eq. (\ref{eq:lightst}),
the trilinear mass term, $\widetilde{A}_t$,
must be $|\widetilde{A}_t|^2 \ll m_Q^2$
in order to avoid the suppression of  the
effective coupling of the lightest stop to the Higgs.
Negative values of the right handed stop mass parameter
open the window for electroweak baryogenesis. However, they
may be associated with the appearence of color breaking minima
at zero and finite temperature. It is hence important to
discuss briefly the constraints on $\widetilde{m}_U$ which
may be obtained from the requirement of avoiding color breaking
minima deeper than the physical one. 

\subsection{Color Breaking Minima}

Let us first analyse the possible existence of color breaking minima 
in the case of zero stop  mixing. In this case,
since $m_Q^2 \gg |m_U|^2$ the only fields which  
may  acquire vacuum
expectation values are the fields $\phi$ and
$U$. At zero temperature, the effective potential is given by
\be
V_{eff}(\phi,U) = -m_{\phi}^2 \phi^2 + \frac{\lambda}{2} \phi^4 + m_U^2 U^2
+ \frac{\widetilde{g}_3^2}{6} U^4 + \widetilde{h}_t^2 \sin^2\beta \phi^2 U^2
\label{effpphiu}
\ee
where $\lambda$ is the radiatively corrected
quartic coupling of the Higgs field,
with its corresponding dependence on the
top/stop spectrum through the one loop
radiative corrections, $\widetilde{g}_3^2/3$ is the radiatively
corrected quartic self-coupling of the field $U$ and
$\widetilde{h}_t^2$ is the bi-bilinear $\phi-U$ coupling.
The latter couplings are well approximated by $\widetilde{g}_3 \simeq g_3$ and
$\widetilde{h}_t \simeq h_t$.
The minimization of this potential leads to three extremes, at: {\bf (i)}
$\phi =0$, $U\neq0$;
{\bf (ii)} $U=0$, $\phi \neq 0$ and {\bf (iii)} $\phi \neq 0$, $U \neq 0$.
The corresponding expressions for the vacuum fields are:
\begin{equation}
\label{solutions}
\begin{array}{rllll}
{\bf (i)}&  U & = & 0, &
{\displaystyle \phi^2 = \frac{m_{\phi}^2}{\lambda}; }
\\ & & & & \\
{\bf (ii)}& \phi & = & 0, &
{\displaystyle U^2 = \frac{3 \widetilde{m}_U^2}{\widetilde{g}_3^2}; }
\\ & & & & \\
{\bf (iii)} & \phi^2 & = &
{\displaystyle \frac{m_{\phi}^2 - 3 \widetilde{m}_U^2 \widetilde{h}_t^2
\sin^2\beta/\widetilde{g}_3^2}{\lambda - 3 \widetilde{h}_t^4 \sin^4\beta/
\widetilde{g}_3^2}, }  &
{\displaystyle U^2 =
\frac{\widetilde{m}_{U}^2 - m_{\phi}^2  \widetilde{h}_t^2
\sin^2\beta/\lambda}{\widetilde{g}_3^2/3 -
\widetilde{h}_t^4 \sin^4\beta/\lambda}. }
\end{array}
\end{equation}
It is easy to show that the branch (iii) is continuosly connected with
branches (i) and (ii). One can also show that the branch (iii) defines
a family of saddle point solutions, the true (local) minima being defined by
(i) and (ii). Hence, the requirement of absence of a color breaking
minimum deeper than the physical one is given by~\cite{CQW}
\begin{equation}
\widetilde{m}_U \leq
\left( \frac{m_h^2 \; v^2 \; \widetilde{g}_3^2}{12} \right)^{1/4}.
\label{boundmu}
\end{equation}
For a  Higgs mass $m_h \simeq 70$ GeV, the bound on
$\widetilde{m}_U$ is of order 80 GeV. This must be compared
with the characteristic value of $\Pi_R(T) \simeq 
{\cal O}\left((100 {\rm GeV})^2\right)$,
implying that even for values of $\tilde{m}_U$ close to the upper
bound on this quantity, a non-negligible screening to the effective
cubic term contributions will be present.

It can also be  shown that, 
for $m_Q \gg \widetilde{m}_U$, the bound $\widetilde{m}_U$
derived above, Eq. (\ref{boundmu}) is sufficient to assure the stability
of the physical ground state for all values of $\widetilde{A}_t$
such that the experimental limits on the lightest stop mass are
preserved.
As we shall show quantitatively below, and it is
clear from our previous discussion,  Eq.~(\ref{totalE}),
large values of $\widetilde{A}_t$
induce a large suppression of the
potential enhancement in the strength of the first
order phase transition through
the light top squark,   and are hence  disfavoured from the point
of view of electroweak baryogenesis.

One must also consider the conditions under which the potential may be 
metastable, but with a lifetime larger than the age of the universe.
Even in this case, in general, the constraint
\be
-\widetilde{m}_U^2+\Pi_R(T_c) > 0
\label{stability}
\ee
must be fulfilled.
Indeed, if Eq.~(\ref{stability}) were not fulfilled, the universe
would be driven to a charge and color breaking minimum
at $T\geq T_c$ (see Eq.(\ref{upot}) below).
 A more conservative requirement can be obtained demanding  that the
 critical temperature for the transition to
the color breaking 
minimum, $T_c^U$, should  be below $T_c$. Due to the strength of
the stop coupling to the gluon and squark fields, one should
expect the color breaking phase  transition to be stronger  than the
electroweak  one.

Let us analyse the finite temperature
effective potential for the $U$ field, which is given by~\cite{CQW}
\begin{equation}
V_U = \left(-\widetilde{m}_U^2 + \gamma_U T^2 \right) U^2 -
T E_U U^3 + \frac{\lambda_U}{2} U^4,
\label{upot}
\end{equation}
where
\begin{eqnarray}
\gamma_U & \equiv & \frac{\Pi_R(T)}{T^2} \simeq
\frac{4 g_3^2}{9} + \frac{h_t^2}{6}\left[ 1 +
\sin^2\beta (1 - \widetilde{A}_t^2/m_Q^2)
\right] ; \;\;\;\;\;\;\;\;\;\;\;\;\;\;
\lambda_U \simeq \frac{g_3^2}{3}
\nonumber\\
E_U & \simeq &  \left[\frac{\sqrt{2} g_3^2}{6 \pi} \left( 1 +
\frac{2}{3\sqrt{3}} \right) \right] 
\nonumber\\
& + &
\left\{ \frac{g_3^3}{12\pi}\left(\frac{5}{3\sqrt{3}} + 1\right)
+ \frac{h_t^3 \sin^3\beta (1-\widetilde{A}_t^2/m_Q^2)^{3/2}}{3 \pi} 
\right\}.
\label{totalEu}
\end{eqnarray}
The
contribution to $E_U$ inside the square brackets comes
from the transverse gluons, $E_U^g$, while the one inside the curly
brackets comes from the squark and Higgs contributions
[Although included in the numerical results,
in the above we have not written explicitly 
the small hypercharge contributions
to $E_U$ and $\gamma_U$.].  We  ignore the gluino and
left handed squark contributions since they are assumed to be heavy and,
hence, their contributions to the finite
temperature effective potential is Boltzman suppressed.
Observe that we have written the contributions that would be obtained if
the field-independent effective thermal mass terms of the
squark and Higgs fields were exactly
vanishing at the temperature $T_c$. Although for values of
$\widetilde{m}_U^2$ which induce a large cubic term in the
Higgs potential, $T_c$ is actually
close to the temperature at which these masses vanish, an
effective screening is always present. This means that
the value of $E_U$ given above is somewhat overestimated.

The difference between $T_0^U$, the temperature at which
$\mstopeff(\phi = 0)  = 0$, and $T_c^U$, is given by
\begin{equation}
T_c^U = \frac{T_0^U}{\sqrt{1 - E_U^2/ 2\lambda_U\gamma_U }}.
\end{equation}
In order to assure a transition from the $SU(2)_L \times U(1)_Y$
symmetric minimum to the physical one at $T = T_c$, we
should replace the condition (\ref{stability}) by the condition
which assures that $T_c^U < T_c$~\cite{CQW},
\begin{equation}
-\widetilde{m}_U^2 + \Pi_R(T) > \widetilde{m}_U^2
\frac{\epsilon}{1-\epsilon} \simeq \widetilde{m}_U^2 \epsilon,
\label{stability2}
\end{equation}
with $\epsilon = E_U^2/2\lambda_U\gamma_U$, a small number.
In the following, we shall require the stability condition,
Eq.~(\ref{stability2}), while using the value of $E_U$
given in Eq.~(\ref{totalEu}). We shall also show the result that
would be obtained if only the gluon contributions to $E_U$,
$E_U^{g}$, would
be considered. The difference between both procedures is just a
reflection of the uncertainties 
involved in our analysis.

\subsubsection{Strength of the First Order Phase Transition 
 in the Large $m_A$ Limit}

Let us first present the results for zero mixing.
Fig.~2 shows the order parameter $v(T_c)/T_c$
for the phase transition as a function of the running light stop mass,
for $\tan\beta = 2$, $m_Q = 500$ GeV
and $M_t = 175$ GeV. These parameters imply a  Higgs mass
$m_h \simeq 70$ GeV, a result which depends weakly on $\widetilde{m}_U$.
We see that for smaller (larger) values of
$\mstop$ ($\widetilde{m}_U$),  $v(T_c)/T_c$
increases in accordance with the above discussion in this section.
The  diamond in fig.2 marks the lower bound on the stop mass coming from
the bound on color breaking vacua at $T=0$, Eq.~(\ref{boundmu}).
The cross and the star denote the bounds that would be obtained
by requiring the condition (\ref{stability2}), while using
the total and gluon-induced trilinear coefficient,
$E_U$ and $E_U^g$, respectively.
We see that the light stop effect is
maximum for values of $\widetilde{m}_U^2$ such that
condition (\ref{stability2}) is saturated,
which leads to values of  $\mstop \simeq 140$ GeV
($\widetilde{m}_U\simeq 90$ GeV) and $v(T_c)/T_c \simeq 1.75$.
\begin{figure}
\vspace{-2cm}
\centerline{
\psfig{figure=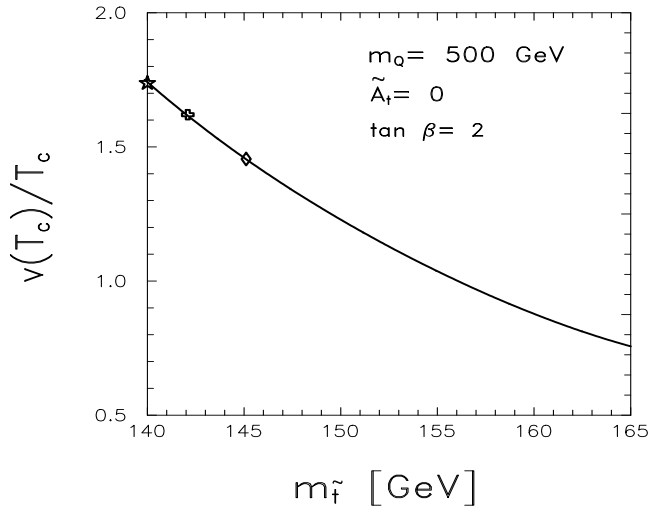,width=12cm,height=10cm}}
\vspace{-1cm}
\caption[]{{\it $v(T_c)/T_c$ as a function of $m_{\tilde{t}}$ for 
$M_t$ = 175 GeV, $m_Q =$ 500 GeV, $\tilde{A}_t$ = 0 and $\tan\beta$ =2.
The diamond [cross, star] denotes the value of $\tilde{m}_U$ for which the 
bound, Eq.~(\ref{boundmu}) is saturated
[ Eq. (\ref{stability2}), while using
the total and gluon induced trilinear coefficients, $E_U$ and $E_U^g$] 
}}
\label{fig:2}
\end{figure}
The preservation of condition
(\ref{boundmu}) demands slightly larger stop mass values. The analysis 
shows that
there is a large region of parameter space for which
$v(T_c)/T_c \geq$~1 and is not in conflict with any phenomenological
constraint.

Fig.~3 shows the results of $v(T_c)/T_c$ for zero mixing and
$m_Q = 500 $ GeV as a function of $\tan\beta$
and for the values of $\widetilde{m}_U$ such that the maximum
effect is achieved.
We also plot in the figure the corresponding values of the stop and Higgs
masses. As in Fig.~2,
the solid [dashed] line
represents the result when the bound  (\ref{boundmu})
[the stability bound of Eq.~(\ref{stability2})] is saturated.
We see that $v(T_c)/T_c$
increases for lower values of $\tan\beta$, a change mainly associated with
the decreasing value of the Higgs mass, or equivalently, of the Higgs
self-coupling. For values of $\tan\beta \simeq 2.7$, 
one gets  $v(T_c)/T_c \simeq 1$,
and hence the value of the Higgs mass yields the upper bound
consistent with electroweak baryogenesis.
This bound is approximately given by $m_h \simeq 80$ GeV. If the bound
on color breaking minima, Eq.~(\ref{boundmu}), is ignored, 
then condition (\ref{stability2}) yields an  upper
bound on $m_h$ close to 100 GeV, in accordance with the 
qualitative discussion presented above (Similar bounds on 
the Higgs mass, $m_h\! < \!100\!$ GeV, 
are obtained when two loops corrections are included and 
condition~(\ref{boundmu}) is saturated \cite{inprep}).
\begin{figure}
\vspace{-2 cm}
\centerline{
\psfig{figure=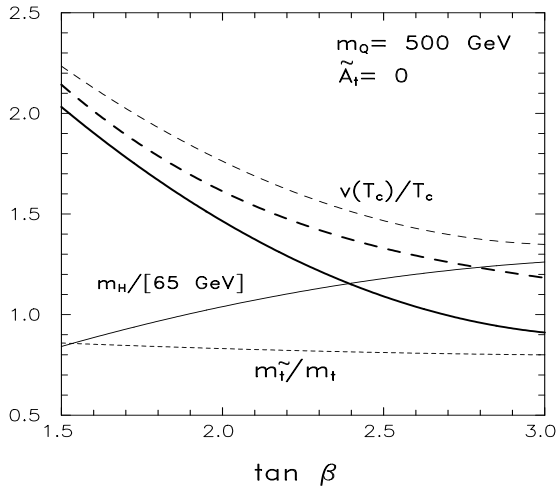,width=12cm,height=10cm}}
\vspace{-1cm}
\caption[]{{\it $v(T_c)/T_c$ as a function of $\tan\beta$  for $m_Q$ and  
$\tilde{A}_t$  as in Fig. 2 and 
$\widetilde{m}_U$ saturating  Eq.~(\ref{boundmu}) [solid]
and  Eq. (\ref{stability2}) [thick dashed line when considering  
the total trilinear coefficient and thin dashed line 
for the gluon-induced part only]. The additional thin lines are plots of
$m_h$ in units of 65 GeV [solid] and 
$\tilde{m}_t/m_t$ [short-dashed line], 
corresponding to the values of $\tilde{m}_U$ associated with the solid line.
}}
\label{fig:3} 
\end{figure}
Due to the logarithmic dependence of $m_h$ on $m_Q$, larger values of
$m_Q$ have the effect of enhancing the Higgs mass values. It turns out
that, for zero mixing, the results for $v(T_c)/T_c$ depend on the Higgs mass
and on the value of $m_U$, but not on the specific value of $m_Q$. Hence,
different values of $m_Q$ have the only effect of shifting (up or down) the
preferred values of $\tan\beta$. In particular,
the fixed point solution, which
corresponds to values of $\tan\beta \simeq 1.6$ for  $M_t$ = 175 GeV,
leads to values of $m_h \simgt 70$ GeV and
$v(T_c)/T_c \simgt 1$, so far $m_Q$ is above 1 TeV and below a
few TeV.

The effect of mixing in the stop sector is very important for the present
 analysis.
For fixed  values of $m_Q$
and $\tan\beta$, increasing values of
$\widetilde{A}_t$ have a negative effect
on the strength of the first order phase transition for three reasons. First,
large values of $\widetilde{A}_t$ lead to larger values of the Higgs mass
$m_h$. Second, as shown in Eq.~(\ref{totalE}) they suppress the stop
enhancement of the cubic term. Finally, there is an indirect effect associated
with the constraints on the allowed values for
$\widetilde{m}_U$. 
This has to do
with the fact that for larger values of
$\widetilde{A}_t$, the phase transition
temperature increases, making more difficult an effective suppression of
the effective mass $\mstopeff$, Eq.~(\ref{plasm}).
Of course, this third reason is absent if the bound (\ref{boundmu})
is ignored.
As we have shown above, for zero mixing the bounds (\ref{orderp}),
(\ref{boundmu})  and  (\ref{stability2})
are only fulfilled for values
of the stop mass larger than approximately 140 GeV.
Light stops, with masses $\mstop \simlt 100$ GeV,
can only be consistent with these constraints for
larger values of the mixing mass parameter $\widetilde{A}_t$.

%
\begin{figure}
\vspace{-2 cm} 
\centerline{
\psfig{figure=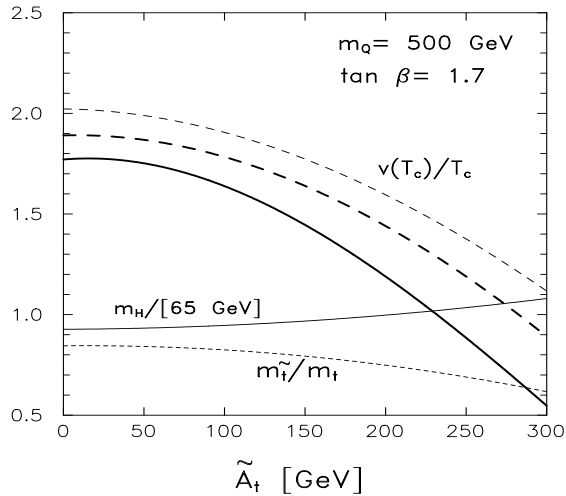,width=12cm,height=10cm}}
\vspace{-1cm}
\caption[]{{\it The same as Fig. 3 but as a function on $\tilde{A}_t$ for
$\tan \beta $ =1.7
}}
\label{fig:4}
\end{figure}

Fig.~4 shows the result for $v(T_c)/T_c$ as a function of $\widetilde{A}_t$
for $\tan\beta = 1.7$,
$m_Q$ = 500 GeV, and values of $m_U$ such that the maximal light
stop effect is achieved.
The same conventions as in Fig.~3 have been used.
Due to the constraints on $\widetilde{m}_U$,
light stops with $\mstop \simlt M_W$, may only be obtained for values of
$\widetilde{A}_t \simgt 0.6\; m_Q$. For these values of $\widetilde{A}_t$,
the phase transition temperature is large
enough to induce large values of $\mstopeff$,
for all values of $\widetilde{m}_U$
allowed by Eq.~(\ref{boundmu}).
In Fig.~4, we have chosen the parameters such that they lead to the
maximum value of the mixing parameter $\widetilde{A}_t/m_Q$ consistent
with $v(T_c)/T_c \geq 1$ and the Higgs mass bound. For values
of $\widetilde{m}_U$ such that
the bounds on color breaking minima are preserved, the mixing
effects on the stop masses are small, and the lightest stop
remains heavier than 100 GeV.
If, however, the weaker bound, Eq.~(\ref{stability2}), were required
(thin and thick dashed lines in Fig.~4), light stops,
with masses of order $80$--$90$ GeV would not be in conflict with
electroweak baryogenesis.

\subsubsection{Sources of CP-violation and the CP-odd Higgs Mass}
\noindent

The new source of CP-violation, beyond the one contained
in the Cabibbo-Kobayashi-Maskawa matrix,
may be either explicit~\cite{ex1} or
spontaneous~\cite{noi} in the Higgs sector (which requires
at least two Higgs doublets). In both cases,
particle mass matrices acquire a
nontrivial space-time dependence when bubbles of the broken
phase
nucleate and expand during a first-order electroweak phase
transition. The crucial observation is that this
space-time dependence cannot be rotated away at two
adjacent points by the same unitary transformation. This provides
sufficiently
fast nonequilibrium CP-violating  effects inside the wall of a
bubble of broken phase expanding in the plasma and may give
rise to a nonvanishing baryon asymmetry through the anomalous
$(B+L)$-violating transitions~\cite{sp} when particles
diffuse to the exterior of the advancing bubble.

As we already mentioned,  new CP-violating phases
arise through the soft supersymmetry breaking parameters associated
with the left-right stop mixing, namely $A_t$ and $\mu$, 
Eq. (\ref{eq:stopmatrix}). The
stop induced current is hence proportional to the variation of
the phase of  $(A_t H_2 - \mu^* H_1) = |A_t H_2 - \mu^* H_1|
\exp(i \phi_{\tilde{A}})$. It is easy to show that 
\begin{equation}
\partial_{\nu} \phi_{\tilde{A}} \sim
{\rm Im}[A_t \mu] \left(H_2 \partial_{\nu}H_1 - H_1
\partial_{\nu}H_2 \right).
\end{equation}

The phase of $\mu$ enters also in the chargino sector. If we
consider the chargino square mass matrix
\begin{equation}
{\cal M}_{\rm ch} {\cal M}_{\rm ch}^{\dagger} =
\left[ \begin{array}{cc}
M_2^2 + g^2 H_2^2 &  g(M_2 H_1 + \mu^* H_2) \\
g(M_2 H_1 + \mu H_2)  & |\mu|^2 + g^2 H_1^2 \end{array}
\right],
\end{equation}
where $g = 2 M_w/v$ is the SU(2) gauge coupling and 
$M_2$ is the (assumed) real soft supersymmetry breaking mass of the
supersymmetric partners of the weak gauge bosons,
the chargino induced
CP-violating current is proportional to the variation
of the phase of the mixing term, $(M_2 H_1 + \mu^* H_2) =
|M_2 H_1 + \mu^* H_2| \exp(i\phi_{\tilde{\mu}})$. It follows
that
\begin{equation}
\partial_{\nu} \phi_{\tilde{\mu}} \sim
{\rm Im}[M_2 \mu] \left(H_2 \partial_{\nu}H_1 - H_1
\partial_{\nu}H_2 \right).
\end{equation}
Defining $H^2 = H_1^2 +H_2^2$,
to a good approximation,
the currents are proportional to
the function 
\begin{equation}
H_1(z) \partial_z H_2(z) - H_2(z) \partial_z H_1(z)
\equiv H^2(z) \partial_z\beta(z),
\label{current}
\end{equation}
with $z$ the time component of the four vector $z_{\nu}$. Since the
time variation of the Higgs fields in the plasma frame
is due to the expansion of the bubble
wall through the thermal bath, ignoring the 
curvature of the bubble wall and
assuming 
that the bubble wall 
is moving along the $z_3$ axis with velocity $v_w$,
any quantity becomes  a  function of
${\bf z}= z_3 + v_w z$, the coordinate normal to the wall surface. 
Eq. (\ref{current})
should vanish smoothly for values of ${\bf z}$ outside the
bubble wall. 
Since $\partial_z\beta \equiv v_w \partial_{\bf z} \beta$ in
Eq. (\ref{current})
denotes the derivative of the ratio of vacuum
expectation values of the Higgs fields, 
it will
be non-vanishing only if the CP-odd Higgs mass takes values of
the order of the critical temperature.

Values of the CP-odd Higgs mass $m_A \simlt 200$ GeV
are, however,  
associated with a weaker first order phase transition. 
Fig.~5
shows the behaviour of the order parameter $v/T$ in the
[$m_A-\tan\beta$] plane, for $\widetilde{A}_t = 0$, $m_Q = 500$
GeV and
values of $\widetilde{m}_U$ close to its upper bound,
Eq.~(\ref{boundmu}). In order to interpret correctly the results
of Fig.~5 one should remember that the Higgs mass bounds are
somewhat
weaker for values of $m_A < 150$ GeV. However, even for values
of $m_A$ of order 80 GeV, in the low $\tan\beta$ regime the lower
bound on the Higgs mass is of order 60 GeV. Hence, it follows
from Fig.~5 that, to obtain a sufficiently strong first order
phase transition, $v(T_c)/T_c \simgt 1$, the CP-odd Higgs mass
must fulfill the condition
$m_A \simgt 150$ GeV.

In order to compute the CP-violating sources, 
the variation of the angle $\beta$ along the bubble wall 
should be computed. The Higgs profiles along the wall
are likely to follow the path of minimal energy connecting
the electroweak symmetry preserving
and the symmetry breaking
vacua in the Higgs potential. The Higgs
potential in the case of low values of the CP-odd Higgs 
mass may be computed by methods similar to those ones explained
in section 2, by preserving the field dependence on both 
Higgs fields \cite{mariano2,CQRVW}. For small values of the fields
$H_i$, as those appearing
close to the symmetric phase,
the Higgs potential for the neutral CP-even components of
the Higgs doublets, $H_1$ and $H_2$,
may be approximated by
\begin{equation}
V(H_1,H_2,T) = m_1^2(T) H_1^2 + m_2^2(T) H_2^2 - 2 m_3^2(T)
H_1 H_2 + {\cal O}(H_i^3) T +...
\end{equation}
The value of $\beta$ close to the symmetric phase may be
easily computed at the temperature $T_0$ at which the curvature
at the origin vanishes \footnote{Strictly speaking, we are
interested in the behaviour of the potential at $T=T_c$.
Given the small quantitative difference between $T_0$
and $T_c$, we shall identify
both temperatures in the following discussion},
\begin{equation}
m_3^4(T_0) = m_1^2(T_0) m_2^2(T_0).
\end{equation}

\begin{figure}
\centerline{
\psfig{figure=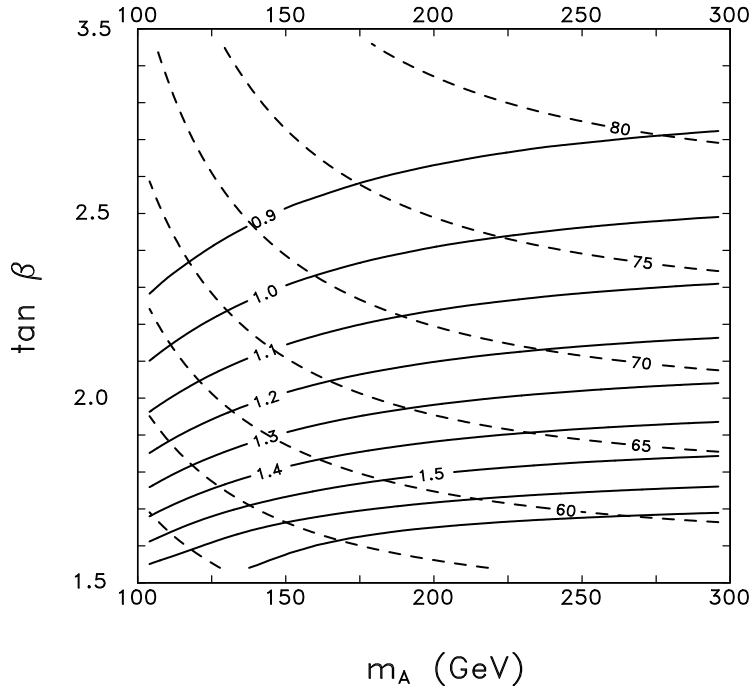,width=8cm,height=9.0cm,bbllx=6.cm,bblly=.cm,bburx=20.5cm,bbury=17cm}}
\vspace{-.3 cm}
\caption{Contour plots of constant values of $v(T_c)/T_c$ (solid
lines) and $m_h$ in GeV (dashed lines) in the plane
$(m_A,\tan\beta)$. We have fixed $M_t=175$ GeV and the values
of sypersymmetric parameters: $m_Q=500$ GeV, 
$\widetilde{m}_U= \widetilde{m}_U^{c}$
 fixed by the charge and color breaking constraint, and
$A_t=\mu^*/\tan\beta$.}
\label{f1}
\end{figure}

Under these conditions, the perturbations of the Higgs fields
close to the origin will follow the path such that
the value of the potential is minimized along it, namely,
\begin{equation}
m_1^2(T_0) v_1^2 + m_2^2(T_0) v_2^2 - 2 m_3^2(T_0) v_1 v_2 = 0
\end{equation}
or, equivalently,
\begin{equation}
\tan^2\beta(T_0,H_1\simeq 0,H_2\simeq 0) = 
\frac{m_1^2(T_0)}{m_2^2(T_0)} \simeq
 \frac{m_1^2(T_c)}{m_2^2(T_c)}.
\end{equation}

The exact value of $\beta$ at the critical temperature may
be  computed by a numerical simulation of the full
effective potential. Hence, the variation of $\beta$ 
along the bubble wall may
be approximately given by
\begin{equation}
\Delta\beta \simeq \beta(T_c) - \arctan(m_1(T_c)/m_2(T_c)).
\end{equation}
This quantity tends to zero
for large values of $m_A$
 like  $\Delta\beta \sim H^2/m_A^2$.
 A numerical estimate \cite{CQRVW}
gives that, for $m_A = 200$ GeV, $\Delta\beta \simeq 0.015$
and hence  values of $m_A > 300$ GeV imply a strong 
suppression of the generated CP-violating sources.
We shall fix
$m_A = 200$ GeV
for most of the following analyis.

\section{GENERATION OF THE BARYON ASYMMETRY}

Baryogenesis is
fueled by CP-violating sources which  are
locally induced by the passage of the wall~\cite{thick,thicknoi}.
These sources do not provide net baryon number. Indeed, in the absence
of baryon number violating processes, the generated baryon 
number will be zero. Since the CP-violation sources are
non-zero only inside the wall, it was first thought that a detail
analysis of the rate of anomalous processes inside the bubble wall
was necessary to estimate the generated baryon number.
However, these first analyses ignored the crucial role played
by diffusion~\cite{tra}.  Indeed,
transport effects  allow  CP-violating charges to  efficiently
diffuse into the symmetric phase 
--in front of the advancing bubble wall-- where anomalous electroweak
baryon number violating processes are unsuppressed.
This amounts to  greatly enhancing the final baryon asymmetry.

In order to estimate the generated baryon number, a set of
coupled
classical Boltzmann equations describing particle distribution
densities should be solved. These equations take into account
particle number changing reactions~\cite{cha} and they allow
to trace the crucial role played by diffusion~\cite{tra}.
Since the weak anomalous processes affect only the
left handed quarks and leptons, the relevant CP-violating 
sources are those which can lead through particle interactions
 to a net chiral charge
for the Standard Model quarks. The new CP-violating
sources we are considering are associated with the parameters
$A_t$ and $\mu$, therefore, the relevant currents are the stop, chargino
and neutralino ones. Although the masses of the first and
second generation squarks, as well as the sbottom ones, are
affected by the  phase of the $\mu$ parameter,
they couple very weakly to the Higgs and hence they play no
role in the computation of the CP-violating currents. 

The CP-violating sources for left- and right-handed squarks, 
charginos and neutralinos are converted into sources 
of chiral quarks via
supergauge and  top quark Yukawa interactions, respectively. Indeed, the
top Yukawa coupling is sufficiently strong, so that the 
top Yukawa induced processes are in approximate thermal equilibrium.
The same happens with the supergauge interactions, if the gauginos
are not much heavier than the critical temperature. Moreover, the
strong sphaleron processes are the most relevant sources of
first and second generation chiral quarks, and hence they must 
be taken into account while computing the generated baryon number.

The stop, chargino and neutralino currents at finite temperature
may be computed by using diagramatic methods~\cite{chou}$^-$\cite{riotto}. 
For small values
of the mixing mass parameter, $|\tilde{A}_t|/m_Q < 0.5$, and large
values of $m_Q \gg T$, the  CP-violating stop induced current is
naturally suppressed. The 
current  associated with neutral and charged higgsinos is the most
relevant one, and it 
may be written as~\cite{CQRVW}

\be
\label{corhiggs}
J^{\mu}_{\widetilde{H}}=\overline{\psi}\gamma^\mu \psi
\ee
where $\psi$ is the Dirac spinor
\be
\label{Dirac}
\psi=\left(
\begin{array}{c}
\widetilde{H}_2 \\
\overline{\widetilde{H}}_1
\end{array}
\right)
\ee
and $\widetilde{H_2}=\widetilde{H}_2^0$ ($\widetilde{H}_2^+$),
$\widetilde{H_1}=\widetilde{H}_1^0$ ($\widetilde{H}_1^-$) for
neutral (charged) higgsinos.

The vacuum expectation value of the (zero component of the)
higgsino
current is approximately given by \footnote{We display here only
the dominant contribution to the current~\cite{CQRVW}}
\begin{equation}
\langle J_{\widetilde{H}}^0(z)\rangle \simeq
{\rm Im}(\mu)\: \left(H_1(z) \partial_z H_2(z) - H_2(z)
\partial_z H_1(z) \right)
\left[ 3 M_2 \; g^2 \; {\cal G}^{\widetilde{W}}_{\widetilde{H}}
\right],
\label{currenth}
\end{equation}
where
\begin{eqnarray}
{\cal G}^{\widetilde{W}}_{\widetilde{H}} & \simeq & 
\int_0^\infty dk
\frac{k^2}
{2 \pi^2 \Gamma_{\widetilde{H}}
\omega_{\widetilde{H}} \omega_{\widetilde{W}}} 
\left( {\rm Im}(n_{\widetilde{H}}) +
{\rm Im}(n_{\widetilde{W}}) \right)
I_2(\omega_{\widetilde{H}},
\Gamma_{\widetilde{H}},
\omega_{\widetilde{W}},\Gamma_{\widetilde{W}})
\phantom{\frac{1}{2^2}} 
\nonumber\\
\end{eqnarray}
with $n_{\widetilde{H}(\widetilde{W})} =
1/\left[\exp\left(\omega_{\widetilde{H}(\widetilde{W})}/T
+ i \Gamma_{\widetilde{H}(\widetilde{W})}/T \right)
+ 1 \right]$, where
$\omega^2_{\widetilde{H}}
=k^2+ |\mu|^2, \;\;$
$\omega^2_{\widetilde{W}}
=k^2+ M_2^2$, 
while $\Gamma_{\widetilde{H}}$ and $\Gamma_{\widetilde{W}}$
are
the damping rate of charged and neutral
Higgsinos and winos, repectively. Since these damping
rates are
dominated by weak interactions \cite{weldon,henning}, 
we shall take
$\Gamma_ {\widetilde{H}} \simeq \Gamma_{\widetilde{W}}$
to be of order of $ 5\times 10^{-2} T$.
Moreover,
the function $I_2$ is given by \cite{CQRVW}
\begin{equation}
I_2(a,b,c,d) = \frac{r_1^2 - 1}{2 \left(r_1^2 + 1 \right)
\left[(a+c)^2 + (b+d)^2 \right]} +
\frac{r_2^2 - 1}{2 \left(r_2^2 + 1 \right)
\left[(a-c)^2 + (b+d)^2 \right]},
\nonumber \\
\;\;
\end{equation}
where $r_1 = (a+c)/(b+d)$ and $r_2 = (a-c)/(b+d)$.
The above expression, Eq. (\ref{currenth}), proceeds from an
expansion in derivatives of the Higgs field and it is valid only
when the mean free path $\Gamma_{\widetilde{W}(\widetilde{H})}^{-1}$ 
is smaller
than the scale of variation of the
Higgs background determined by the wall thickness and the wall
velocity, $\Gamma_{\widetilde{W}(\widetilde{H})} L_w/v_w \gg 1$.

As mentioned before, the above currents may be used to
compute the particle densities, once diffusion and particle
changing interaction effects are taken into account. We shall
not discuss this in detail here, but we shall limit ourselves to
present the most important aspects related to the generation of
baryon number.

The particle densities we
need to include are the left-handed top
doublet $q_L\equiv(t_L+b_L)$,
the right-handed top quark $t_R$, the Higgs particle
$h\equiv(H_1^0, H_2^0, H_1^-, H_2^+)$, and the superpartners
$\widetilde{q}_L$, $\widetilde{t}_R$ and $\widetilde{h}$.
The interactions able to change the particle numbers are the top
Yukawa interaction with rate
$\Gamma_t$, the top quark mass interaction with rate $\Gamma_m$,
the Higgs self-interactions in the broken phase
with rate $\Gamma_{\cal H}$, the strong sphaleron interactions
with rate $\Gamma_{{\rm ss}}$,
the weak anomalous interactions with rate $\Gamma_{\rm ws}$
and the gauge interactions.
The system may be described by the
densities
${\cal Q} = q_L + \widetilde{q}_L$,
${\cal {\cal T}}=t_R+\widetilde{t}_R$ and ${\cal H}=h+\widetilde{h}$.
CP-violating interactions with the advancing bubble wall produce
source terms $\gamma_{\widetilde{H}}=\partial_z\langle
J_{\widetilde{H}}
^0(z)\rangle$ for Higgsinos and
$\gamma_R=\partial_z \langle J_{R}^0(z)\rangle$
for right-handed stops, which tend to push the system out of
equilibrium.

If the system is near thermal equilibrium and particles interact
weakly, the particle number densities $n_i$ may be expressed as
$n_i = k_i \mu_i T^2/6$, where $\mu_i$ is
the local chemical potential and $k_i$ are statistical factors of
order 2 (1) for light bosons (fermions) in thermal equilibrium,
and Boltzmann suppressed for particles heavier than $T$.
Assuming that the rates $\Gamma_t$ and $\Gamma_{{\rm ss}}$ are
fast so that ${\cal Q}/k_q-
{\cal H}/k_{\cal H}-{\cal T}/k_{\cal T}={\cal O}(1/\Gamma_t)$ and
$2{\cal Q}/k_q-{\cal T}/k_{\cal T}+
9({\cal Q}+{\cal T})/k_b={\cal O}(1/\Gamma_{{\rm ss}})$,
one can find the equation governing the Higgs density~\cite{newmethod2}
\begin{equation}
\label{equation}
v_{\omega}{\cal H}^\prime-\overline{D}
{\cal H}^{\prime\prime}+\overline{\Gamma}{\cal H}-
\widetilde{\gamma}=0,
\end{equation}
where the derivatives are now with respect to ${\bf z}$,
$\overline{D}$ is the effective diffusion constant,
$\widetilde{\gamma}$ is an effective source term
in the frame of the
bubble wall and $\overline{\Gamma}$ is the effective decay
constant.
An analytical solution to Eq.~(\ref{equation}) satisfying the
boundary conditions ${\cal H}(\pm\infty)=0$ may be found in the
symmetric
phase (defined by ${\bf z}<0$) using a ${\bf z}$-independent
effective diffusion constant 
and a step function for the effective decay rate
$\overline{\Gamma}= \widetilde{\Gamma} \theta({\bf z})$. A more realistic
form of $\overline{\Gamma}$ would interpolate smoothly between the
symmetric and the broken phase values. We have checked, however,
that the result is insensitive to the specific position of the
step function  inside the bubble wall.

The analytical solution to the
diffusion equations for  ${\bf z} < 0$ leads 
to~\cite{CQRVW,newmethod2}
\begin{equation}
\label{higgs1}
{\cal H}({\bf z})={\cal A}\:{\rm e}^{{\bf z}v_{\omega}/\overline{D}},
\end{equation}
and for ${\bf z} >0$,
\begin{eqnarray}
\label{higgs3}
{\cal H}({\bf z}) & = & \left( {\cal B}_{+} -
\frac{1}{\overline{D}(\lambda_+  - \lambda_-)}
\int_0^{{\bf z}} du \widetilde \gamma(u) e^{-\lambda_+ u} \right)
e^{\lambda_{+} {\bf z}}
\nonumber\\
&+& \left( {\cal B}_{-} -
\frac{1}{\overline{D}(\lambda_-  - \lambda_+)}
\int_0^{\bf{z}} du \widetilde \gamma(u) e^{-\lambda_- u} \right)
e^{\lambda_{-} {\bf z}}.
\end{eqnarray}
where
\begin{equation}
\lambda_{\pm} = \frac{ v_{\omega} \pm
\sqrt{v_{\omega}^2 + 4 \widetilde{\Gamma}
\overline{D}}}{2 \overline{D}},
\end{equation}
and $\widetilde \gamma({\bf z}) = v_{\omega} \partial_{{\bf z}}
J_0({\bf z}) f(k_i)$,
$J_0$ being the total CP-violating current resulting
from the sum of the right-handed
stop and Higgsino contributions and
$f(k_i)$  a coefficient depending on the number of
degrees of freedom present in
the thermal bath and related to the definition of the
effective source~\cite{newmethod2}.
Imposing the continuity of ${\cal{H}}$ and
${\cal{H}}'$ at the boundaries, we find~\cite{CQRVW}
\begin{equation}
\label{higgs2}
{\cal A}= {\cal B}_{+}\left(1-\frac{\lambda_-}{\lambda_+}\right)=
{\cal B}_{-}\left(\frac{\lambda_+}{\lambda_-}-1\right)=
\frac{1}{\overline{D} \; \lambda_{+}} \int_0^{\infty} du\;
\widetilde \gamma(u)
e^{-\lambda_+ u}.
\end{equation}
{} From the form of the above equations one can see that CP-violating
densities diffuse in a time $t\sim \overline{D}/ v_{\omega}^2$
and the assumptions leading to the analytical
form of ${\cal H}({\bf z})$ are valid
provided $\Gamma_t,\Gamma_{{\rm ss}}\gg
v_{\omega}^2/\overline{D}$.

The equation governing
the baryon asymmetry $n_B$ is given by~\cite{newmethod2}
\begin{equation}
\label{bau}
D_q n_B^{\prime\prime}-v_{\omega} n_B^\prime-
\theta(-{\bf z})n_f\Gamma_{{\rm ws}}n_L=0,
\end{equation}
where $\Gamma_{{\rm ws}}=6\kappa\alpha_w^4T$ is the
weak sphaleron
rate ($\kappa\simeq 1$)~\footnote{The value
of $\kappa$ is still subject of debate~\cite{SphalRate,ASY} },
and $n_L$ is the total number density of
left-handed weak doublet fermions, $n_f=3$ is the number of
families and
we have assumed that the baryon asymmetry gets
produced only in the symmetric phase.
Expressing $n_L({\bf z})$ in terms of the Higgs 
number density~\cite{newmethod2}
\begin{equation}
n_L=\frac{9k_q k_{\cal T}-8k_b k_{\cal T}
-5 k_b k_q}{k_{\cal H}(k_b+9 k_q+9 k_{\cal T})}\:{\cal H}
\end{equation}
and making use of Eqs.~(\ref{higgs1})-(\ref{bau}), we find that
\begin{equation}
\frac{n_B}{s}=-g(k_i)\frac{{\cal A}\overline{D}\Gamma_{{\rm ws}}}
{v_{\omega}^2 s},
\end{equation}
where $s=2\pi^2 g_{*s}T^3/45$ is the entropy density ($g_{*s}$
being
the effective number of relativistic degrees of freedom) and
$g(k_i)$
is a numerical coefficient depending upon the light degrees of
freedom present in the thermal bath.

\begin{figure}
\vspace{-.6 cm}
\centerline{
\psfig{figure=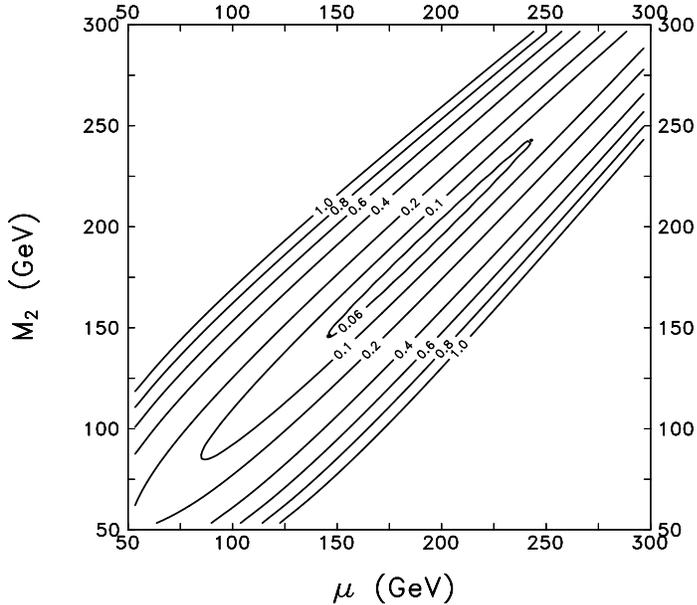,width=9cm,height=8.0cm,bbllx=2.cm,bblly=.cm,bburx=20.5cm,bbury=17cm}}
\caption{Contour plot of $|\sin \phi_{\mu}|$
in the plane ($\mu,M_2$) for fixed $n_B/s = 4 \times 10^{-11}$ and
 $v_{\omega}=0.1$, $L_{\omega}=25/T$, $m_Q=500$ GeV,
$m_U=m_U^{\rm crit}$, $\tan\beta=2$ and
$A_t = \mu^*/\tan\beta$.}
\label{f6}
\vspace{-0.5 cm}
\end{figure}
Fig. 6  shows the value of the phase of the parameter $\mu$ 
needed to obtain the observed baryon asymmetry, 
$n_B/s \simeq 4 \times 10^{-11}$, within the approximations
given above and using a semirealistic approximation for the Higgs
profiles \cite{CQRVW}. 
 The wall velocity is
taken to be $v_{\omega} = 0.1$, while the bubble wall width is
taken to be $L_{\omega} = 25/T$.
Our results, are, however, quite  insensitive to
the specific choice of $v_{\omega}$ and $L_{\omega}$.
It is interesting to note that realistic values of the baryon
asymmetry may only be obtained for values of the CP-violating
phases of order one, and for a very specific region of the 
[$\mu-M_2$] plane. 
Values of the phases lower than 0.1 are only consistent with
the observed baryon asymmetry for values of $|\mu|$ of
order of the gaugino mass parameters. This is due to a 
resonant behaviour of the induced 
Higgsino current for 
$|\mu| \simeq M_2$ (See Eq. (\ref{currenth})). 

In conclusion, the requirement of a sufficiently strong 
first order phase transition and of sizeable CP-violating currents
may be only satisfied within a very specific region of parameters
within the MSSM. The realization of this scenario will imply
very specific signatures which may be tested in  future
runs of existing experimental facilities. We shall expand on this
issue in the next section.

\section{EXPERIMENTAL TESTS OF ELECTROWEAK \\
 BARYOGENESIS}

In the previous sections, we have shown that the scenario
of electroweak baryogenesis favors  Higgs
masses $m_h \simlt 80$ GeV. Slightly heavier Higgs bosons
may  be consistent with this scenario only if higher-order
(or non-perturbative) effects render the phase transition more
strongly first order than what is suggested by one-loop analyses.
A hint in this direction was obtained in  recent works~\cite{JoseR,JRBC}, 
where it was shown that, when two-loop corrections are
included, the requirement of preservation of the generated
baryon asymmetry may be fulfilled, for values of the
Higgs masses of order 80 GeV even for $m_U \simeq 0$ (see, for
comparison, the results of Fig. 2, for which $m_h \simeq 70$
GeV). An ongoing two-loop analysis 
for $m_U^2 \simlt 0$  shows
an interesting extension of the allowed $m_h$ region~\cite{inprep}. 
As we discussed above, the phase transition may also
become moderately 
stronger assuming that the physical ground state is metastable.
In view of all present studies,  it may be concluded that
a Higgs mass above 95 GeV will  put very strong 
constraints on the
scenario of electroweak baryogenesis within the MSSM.
Hence, the most direct experimental way of testing this scenario is
through the search for the Higgs boson at LEP2.

At LEP2, the ligthest CP-even
neutral Higgs bosons may be  produced in association with  $Z$
via Higgs-strahlung
\begin{equation}
e^+ e^- \rightarrow Z^* \rightarrow Z \; h,
\label{eq:rateh}
\end{equation}
or in association with the neutral CP-odd Higgs scalar,
\begin{equation}
e^+ e^- \rightarrow Z^* \rightarrow h \; A.
\end{equation}
The associated $Ah$ production becomes increasingly important for rising
values of $\tan\beta$. However, for values
of the CP-odd Higgs mass above 100 GeV  it is kinematically forbidden,
 and hence, it is not relevant for
testing the scenario of electroweak baryogenesis, for which
$m_A \simgt 150$ GeV.

The Higgs production rate (\ref{eq:rateh}) is equal to the Standard
Model one times a projection factor. This projection factor, takes
into account the component of the lightest CP-even Higgs on the
Higgs which acquires vacuum expectation value (which is the one
which couples to the $Z$ in the standard way),
\begin{equation}
\sigma_{MSSM}(e^+e^- \rightarrow Z \;h) =
\sigma_{SM}(e^+e^- \rightarrow Z \;h) \times
\sin^2(\beta - \alpha).
\end{equation}
For large values of the CP-odd Higgs mass, the heavy Higgs
doublet decouples and $\sin(\beta - \alpha) \rightarrow 1$.
Indeed, the Higgs sector of the theory behaves effectively as
the Standard Model one. This transition is achieved rather
fast, and for CP-odd Higgs masses above 150 GeV, the cross
section differs only slightly from the Standard Model one.
Hence, in the limit of interest for 
this discussion, the mass reach for the lightest CP-even
Higgs boson within the MSSM is almost
indistinguishable from the one of the Standard Model Higgs.

The search for the Higgs boson is performed by taking into account the
dominant decay modes of the Higgs into bottom and $\tau$ pairs. 
Barring the possibility of supersymmetric
decay channels, which only appear in very limited regions of
parameter space, which will be directly tested through
SUSY particle searches, the Higgs decays approximately 90~$\%$ of the
time into $b\bar{b}$ pairs and 8~$\%$ of the time into $\tau
\bar{\tau}$ pairs. The $Z$ boson may decay in jets (70 $\%$),
charged leptons (10 $\%$) or neutrinos (20 $\%$).

\begin{figure}
\vspace{-6 cm}
\centerline{
\psfig{figure=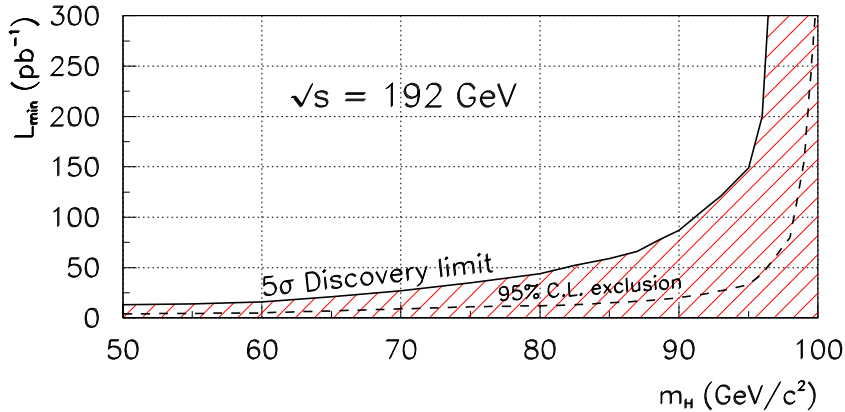,width=12cm,height=12cm}}
\caption[]{{\it
Minimum luminosity needed per experiment, in $pb^{-1}$,
for the combined $5\sigma$ discovery (full line) or the combined 95\% C.L.
exclusion (dashed line) of the Higgs boson as a function of its mass,
at a centre-of-mass energy of $\sqrt{s} = 192$ GeV.}}
\label{fig:lum}
\end{figure}

Based on the experimental simulations \cite{LEPRep}, 
it is possible to derive the exclusion
and discovery  limits for the 
lightest CP-even Higgs  mass 
as a function of the luminosity for the expected 
LEP2 energy range~\footnote{As explained above, we are considering scenarios
with relatively heavy CP-odd Higgs bosons, for which the limits
for the lightest CP-even Higgs coincide with the ones for the SM Higgs}.
The contours are defined
at 5$\sigma$ for the discovery
and 95$\%$ C.L. for the exclusion limits.
The results of the combination of the four experiments
for a center of mass energy of $\sqrt{s} = 192$ GeV
are shown in Fig.~\ref{fig:lum}.  
In Table~\ref{tab:SM} we summarize
the minimum luminosities
which are
needed per experiment
for exclusion and discovery for the largest Higgs mass values
that can be realistically
reached at   center-of-mass energies of 175, 192 and 205 GeV; beyond these
maximum mass values the required luminosities increase sharply for the
exclusion and discovery of the Higgs particle.

\begin{table}
\caption
{{\it Maximal Higgs masses that can be excluded or discovered with
an integrated luminosity $L_{min}$ per experiment at the three
representative energy values of 175, 192 and 205 GeV, if the
four LEP experiments are combined.}}
\label{tab:SM}
\begin{center}
\begin{tabular}{||c||c|c||c|c||}
\hline
   &   &    &  & \\
  &   & Exclusion &  &  Discovery   \\
   &   &   &   &  \\
 $\sqrt{s}$[GeV]& $m_H$[GeV]&$L_{min}$[pb$^{-1}$] & $m_H$[GeV]& $L_{min}$
 [pb$^{-1}$] \\
   &   &  per experiment        & & per experiment \\
   &  &  &  &   \\
\hline
\hline
 &   &   &   & \\
 175 & 83 & 75  & 82 & 150 \\
&   &   & &   \\
\hline
 &   &   &  &  \\
 192 & 98 &  150  & 95 & 150 \\

&   &   & &   \\
\hline
 &   &   &  &  \\
 205 & 112 & 200  & 108 & 300  \\
&   &   &  &  \\
\hline
\end{tabular}
\end{center}
\end{table}

It is important to compare the above  results with the 
currently expected
energy range and luminosity of the LEP2 experiment.
LEP2 is expected to run at a center of mass energy
of $\sqrt{s} \simeq 184$ GeV during the summer of
1997 and to collect a total integrated luminosity
of approximately 100 pb$^{-1}$ per experiment.
Taken into account the results of the
above analysis, LEP2 at $\sqrt{s} \simeq $ 184 GeV 
will be able to discover a 
Standard Model-like Higgs boson with a mass up to
approximately 85 GeV. In case of negative searches, 
this will set a lower
bound on the lightest CP-even Higgs mass of order 90 GeV
for values of $m_A > 150$ GeV.
Clearly, the exact range will
finally depend on the real performance of the experiments.

Therefore, if  the scenario of electroweak baryogenesis
is realized in nature, the 1997 run of the LEP2 experiment
has excellent prospects for detecting a Higgs. If, however,
no signal is found, this will pose very strong constraints on
the present scenario. For instance, in order to preserve a
sufficiently strong first order phase transition
the model will be
driven into a corner of $\tilde{A}_t\ll m_Q $. In
addition, the enhancement of the phase transition due
to higher order effects will be crucial. It is 
clear that non-perturbative information, as well as
a deeper insight on the question of metastability of
the physical vacuum will then be necessary to decide the fate
of this scenario.

Moreover,
the LEP2 experiment is expected to achieve a final center
of mass energy of about $\sqrt{s} \simeq 192$ GeV and to
collect a total integrated luminosity
$L \simgt 100$ pb$^{-1}$ per year and per experiment.
As can be inferred from Table 1,
this will lead to a  discovery limit of order 95 GeV and
an exclusion limit of about  100 GeV. This will 
definitely test  the possibility of baryogenesis at the 
electroweak scale
within the MSSM, since larger values of the Higgs mass
are unlikely to be consistent with this scenario.

If the Higgs is found at LEP2, the second
test will come from the search for the lightest stop at the Tevatron
collider (the stop mass is typically too large for this particle to be seen
at LEP). The stop can be pair produced at the Tevatron through 
gluon processes. It  can subsequently decay into 
bottom and chargino
with almost one hundred percent branching ratio, unless the
chargino mass is very close or above the stop mass.
If this is the  
case, the  stops decays  through a loop into charm and neutralino.
The signal from the tree level decay can be either a single lepton 
plus missing energy and b- and light quark-jets,
 or dilepton plus missing energy 
and b-jets~\footnote{we are 
only considering the case of exact 
R-Parity conservation}, 
\begin{eqnarray}
\tilde{t}  \rightarrow  b \tilde{\chi}^{\pm} \;\;&
\tilde{\chi}^{\pm} \rightarrow  \tilde{\chi}^0_1 \; l^{\pm} \; \nu \;\;&
\tilde{\chi}^{\pm} \rightarrow \tilde{\chi}^0_1 \; qq 
\nonumber \\
\tilde{t}  \rightarrow  b \tilde{\chi}^{\pm} \;\;&
\tilde{\chi}^{\pm} \rightarrow  \tilde{\chi}^0_1 \; l^{\pm} \; \nu \;\;&
\tilde{\chi}^{\pm} \rightarrow \tilde{\chi}^0_1 \; l^{\pm} \; \nu .
\end{eqnarray}
If the 
above channel is kinematically forbidden, the stop 
signal will then be missing energy plus  two acollinear jets,
\begin{equation}
\tilde{t} \rightarrow c \tilde{\chi}^0_1 .
\end{equation}
At present, the D0 experiment has analysed only 
14 pb$^{-1}$ of the 100 pb$^{-1}$ data in the 
$\tilde{t} \rightarrow \tilde{\chi}^0_1 c$ channel and
they are able to search for stop masses up to about 100 GeV, depending on
the values of the neutralino mass considered.
Studies about the prospects for 
stop searches at the Run II of the Tevatron~\cite{TeV2000}
(main injector phase at 2 TeV center of mass energy and 2 $fb^{-1}$ of 
intergrated luminosity ) show a maximal mass reach for stops of 
about 150 GeV in the $\tilde{t} \rightarrow \tilde{\chi}^{\pm}_1 b$ 
channel and about 120 GeV in the $\tilde{t} \rightarrow \tilde{\chi}^0_1 c$ 
channel. Forseen upgrades of the Tevatron achieving a
total integrated luminosity of 10/25 $fb^{-1}$  will allow to discover
a top squark with mass below the top quark, although optimization in the
event selection procedure is necessary, specially in the neutralino-charm 
decay channel. Hence,
already the Run II of the Tevatron to begin in 1999, will start
testing an important region of the  stop mass range consistent
with electroweak baryogenesis. The forseen upgrades, if approved, 
will provide a crucial test of the framework under analysis.

\begin{figure}
\centerline{
\psfig{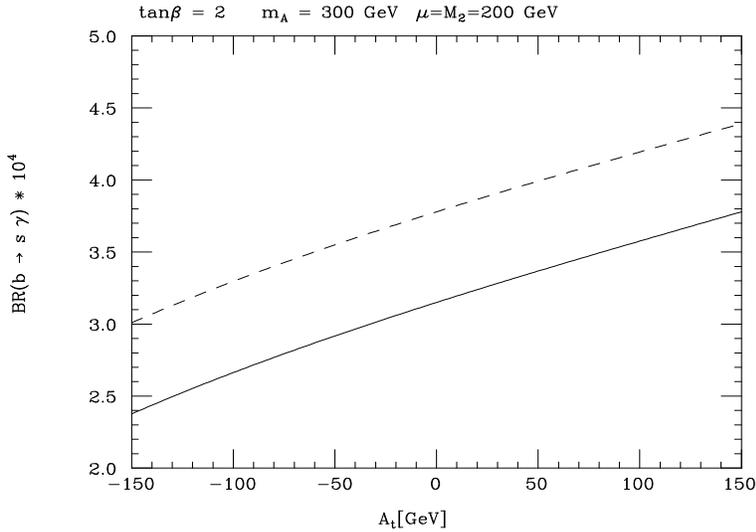}}
\caption{BR($b \rightarrow s \gamma$) as a function of $A_t$,
for $\mu = M_2$= 200 GeV, $m_A$ = 300 GeV, $\tan\beta = 2$, 
$M_t=175$ GeV, $m_Q=500$ GeV, $m_U=m_U^{\rm crit}$
(fixed by the charge and color breaking constraint), and
the first and second generation squark masses equal to 1 TeV.}
\label{bsga}
\end{figure}
If both particles are found, the last crucial test will come
from
$B$ physics. The selected
parameter space leads
to values of the branching ratio ${\rm BR}(b\rightarrow s\gamma)$
different from the Standard Model case \cite{bsgasusy}. 
Although the exact value
of this branching ratio depends strongly on the value of 
$m_A$ and the $\mu$
and $A_t$ parameters, the typical difference with respect to the
Standard
Model prediction is of the order of the present experimental
sensitivity
and hence in principle testable in the near future. Indeed, for the
typical spectrum considered here, due to the 
relatively low values of the light charged Higgs mass,
the
branching ratio ${\rm BR}(b \rightarrow s \gamma)$ is somewhat
higher than in the SM case, unless it is properly cancelled by
the light stop contributions. Figure 8 shows the dependence of
 ${\rm BR}(b\rightarrow s \gamma)$ on $A_t$ for 
$M_2 = \mu = 200$ GeV,
$\widetilde{m}_U = \widetilde{m}_U^c$, $\tan\beta = 2$, $m_Q = 500$
GeV and $m_A = 300$ GeV \footnote{We have ignored
the effect of the CP-violating phases for these
computations}. The solid line represents the 
leading order result
obtained by setting a renormalization scale $Q = m_b$ in the leading
order QCD corrections,  where $m_b$ is the bottom mass. The
dashed line represents the result obtained by setting a normalization
scale $Q = 0.5 \; m_b$, which leads
to results in agreement with the most recent next to leading order
corrections in the Standard Model \cite{nolbsga}.
It is clear from the figure that 
negative values of Re($A_t \times \mu$)  are favored to get consistency
with the present experimental range~\cite{bsgaexp}, 
BR$(b\rightarrow s\gamma)^{\rm 
exp} = (2.3 \pm 0.6) \times 10^{-4}$. 
Since negative values of Re($A_t\times \mu$) 
imply non-negligible mixing in the stop sector, this
rare b-decay imposes very strong constraints on the scenario
of electroweak baryogenesis. 
More information may be obtained by the precise measurement 
of CP-violating asymmetries at  B-factories.
Indeed,
since the light stop couples via superweak interactions
to the bottom sector, the large CP-odd phases associated with
this scenario will naturally imply 
a departure from the Standard Model predictions
for these CP-violating asymmetries .

\section{CONCLUDING REMARKS}

If the scenario of electroweak baryogenesis
is realized in nature, it will demand new physics at scales of
order of the weak scale. A light Higgs, with mass at the reach
of LEP2 will strongly favor this scenario, while new light scalars
with relevant couplings to the Higgs field must also be present. 
These properties are naturally fulfilled within supersymmetric
extensions of the Standard Model. The Higgs is naturally light,
while the new scalars are provided by the supersymmetric partners
of the top quark. Since the stops are charged and colored particles,
their large multiplicity helps in enhancing the strength of the
first order phase transition, allowing the preservation of the
generated baryon number.

The most relevant new
CP-violating sources are associated with the
supersymmetric partners of the charged and neutral Higgs and
weak gauge bosons. These CP-violating sources, which appear through the
Higgsino-gaugino mixing terms, must be of order one in order
to have a relevant effect in the generation of the baryon asymmetry.
Since sizeable phases in the Higgsino mass parameter
could lead to unacceptable values for the 
electric dipole moment of the neutron, one needs to require that
the first and second generation squark masses are of the order of
a few TeV, or else, an unnatural cancellation between different contributions
must take place.

It is interesting to emphazise that the mechanism of electroweak 
baryogenesis can be consistent  with the general framework of unification of
couplings. 
In fact, performing a detailed renormalization group analysis, one can match 
 the specific hierarchy of soft supersymmetry breaking
terms at low energies required by the electroweak baryogenesis
scenario together with large, perturbative values of the top 
Yukawa coupling, as those
associated with the unification of bottom-tau Yukawa couplings or the 
top quark infrared fixed point structure.
A recent study~\cite{CCOPW} shows  that, depending on the scale at which
supersymmetry breakdown is transmitted to the observable sector, the above
implies very specific constraints on the  stop and Higgs mass parameters 
of the theory at  high energies.

Most important, the electroweak baryogenesis explanation 
can be explicitly tested at present and near future experiments. 
In the last phase of LEP2, a center of mass energy $\sqrt{s} 
\simeq 192$ GeV will be achieved, with a total integrated luminosity
of about 100$-$150 pb$^{-1}$ per year and per experiment. 
A lightest CP-even Higgs, with 
mass of about 100 GeV  
is expected to be detectable (or otherwise
excluded) providing a definite test of the scenario of EWB within
the MSSM. 
The CP-odd mass within this framework must be
sufficiently large in order to avoid weakening the first order phase
transition, and it must be sufficiently small to avoid the suppression
of the new CP-violating sources. Altogether this implies that the 
lightest Higgs should be quite Standard model like.
If the Higgs is found, the next test of this scenario 
will come from stop searches at
a high luminosity Tevatron facility. 
Moreover, although this is not required, the charginos and neutralinos
might be light, at the reach of LEP2. 

Another potential experimental test of this model comes from
the rate of flavor changing neutral current processes, like 
$b \rightarrow s \gamma$. We have shown that, if the CP-odd Higgs
mass is not sufficiently
large, this rate will be in general above the
Standard Model predictions, unless Re$(A_t \times\mu) \simlt 0$.
Hence, rare processes put additional constraints on the allowed
parameter space. Moreover, light stops and light charginos, with
additional CP-violating phases associated with the $\mu$ and
$A_t$ parameters, will have a relevant impact on B-physics, which
may be testable at B-factories~\cite{bfac}.

In summary, the realization of electroweak baryogenesis will
not only provide the answer to one of the most interesting open
questions of particle physics, but it will imply a rich 
phenomenology at  present and near future colliders.\\

\section*{Acknowledgements}

We would  like to thank J.R. Espinosa, A. Riotto, I. Vilja and,
in particular,
M. Quiros  for  enjoyable and fruitful 
collaborations related to this subject. We would also like to thank 
 P. Chankowski, J. Cline, 
H. Haber, K. Kainulainen, G. Kane, M. Laine, A. Nelson, M. Pietroni, 
S. Pokorski and M. Shaposhnikov 
for many pleasant and interesting discussions.

\newpage

\end{document}